# Folding Proteins with Both Alpha and Beta Structures in a Reduced Model

**Nan-Yow Chen**

Department of Physics, National Tsing Hua University, Hsinchu,
Taiwan, Republic of China

**Advisor:  Chung-Yu Mou**

Department of Physics, National Tsing Hua University, Hsinchu,
Taiwan, Republic of China

**Advisor:  Zheng-Yao Su**

National Center for High-Performance Computing, Hsinchu,
Taiwan, Republic of China

June, 2004

# Contents

# Abstract

A reduced model, which can fold both helix and sheet structures, is proposed to study the problem of protein folding. The goal of this model is to find an unbiased effective potential that has included the effects of water and at the same time can predict the three dimensional structure of a protein with a given sequence in reasonable time. For this purpose, rather than focusing on the real folding dynamics or full structural details at the atomic scale, we adopt the Monte Carlo method and the coarse-grained representation of the protein in which both side-chains and the backbones are replaced by suitable geometrical objects in consistent with the known structure. On top of the coarse-grained representation, our effective potential can be developed. Two new interactions, *the dipole-dipole interactions* and *the local hydrophobic interactions*, are introduced and are shown to be as crucial as the hydrogen bonds for forming the secondary structures. In particular, for the first time, we demonstrate that the resulting reduced model can successfully fold proteins with both helix and sheet structures without using any biased potential. Further analyses show that this model can also fold other proteins in reasonable accuracy and thus provides a promising starting point for the problem of protein folding.

# Chapter 1

# Introduction

Proteins play important roles in living cells. They provide enormous varieties of biological functions. For instance, acting as enzymes, they catalyze chemical reactions occurring in biological systems and increase rates of the reactions at least by a factor of $10^6$. Proteins can also transport particles ranging from electrons to macromolecules. Haemoglobin is one example, and it transports oxygen and irons in the circulatory system. When residing on the cell membrane, such as Na+-K+-ATPase [1], proteins may form ions pump to keep the concentration balance of ions in cells. In addition, there are huge amounts of hormones, which are also proteins. In this case, proteins act as messengers and are responsible for coordinating cells in tissues. Proteins also help in reading out genetic information stored in DNA. They are synthesized in appropriate quantities at the right moment when translating sequences of DNA. In the immune system, proteins are also indispensable. For instance, proteins can bind to specific foreign invaders such as bacteria or viruses to prevent them to attack the living cells. Finally, proteins are also the building blocks for biological structures. For examples, actin filaments and myosin filaments are the major component of muscles tissues that determine the shape of cells. Clearly, all the above-mentioned functions are essential for our lives, and they cannot work without proteins. A unique feature when proteins function is the high *specificity*. This is reflected not only in selecting molecules they can interact with but also the timing when it happens. In particular, the specificity regarding the molecule proteins interact with is often determined by the steric conformations of proteins. It is almost true that for each biological function, there is an associate protein conformation that can perform the duty. This clearly indicates that the biological functions of proteins are closely related to their three-dimensional structures. Therefore, knowledge of proteins' three-dimensional structures is a necessary step towards understanding how they function.

The high *specificity* implies that given so many biological functions to perform, there cannot be just a few types of proteins. Indeed, designed by Nature, proteins are linear polymers built on combinations of 20 different amino acids, which then allows huge of number of ways for constructing proteins. When proteins are synthesized, they can spontaneously fold into specific three-dimensional structures in aqueous solution to achieve their functions. It is believed that all of the information necessary for folding the protein is contained in the sequence of amino acids [2]. Furthermore, the folding process of protein must satisfy thermodynamic and kinetic conditions. The thermodynamic condition is that the proteins adopt a single, stable, three-dimensional conformation. The kinetic condition is that the protein must fold into its native state in a reasonable timescale. It has been suggested that for a protein with 100 amino acids, a random conformational search would take $10^{31}$ years approximately [3], but proteins fold from milliseconds to seconds. Hence, this paradox of how proteins fold rapidly, as first suggested by Levinthal in 1968 (and hence will be referred as the Levinthal's paradox) and how to determine the three-dimensional structure of proteins from its amino acids sequence are referred to as the *protein-folding problem*.

The existence of kinetic folding pathways [3,4] and folding funnel theory [5-7] are suggested to explain the Levinthal's paradox. As for how proteins acquiring their unique three-dimensional conformation, recent interest revives from studies on the lattice models in which several important insights for protein structure and folding kinetics are obtained [8-10]. Due to their oversimplified nature, insights obtained mainly concern the global picture of folding and are far away from real applications. On the other hand, full-atom investigation would require setting protein in great details including the surrounding water molecules. The number of degrees of freedom for including these water molecules sometimes exceeds that needed for specifying the protein itself. As a result, even the simulation for small size of protein is not feasible under current architectures of computers. Recently, Duan and Kollman performed a full atom simulation of a protein with 36 amino acids [11,12]. Only two trajectories of $1\,\mu s$ duration were obtained by using a powerful parallel computer in about 4 months. Clearly, it demonstrates the non-feasibility of studying the protein-folding problem using the current computer architecture. As a result, much effort has thus been devoted to seek the possibility of solving the protein-folding problem in the so-called reduced models. The challenge is to find a coarse-grained model, without including water molecules explicitly, between the minimalist level (such as the lattice model) and the full-atom level. There are many proposed models in the literature. For examples, there are off-lattice models [13-18] using the Gō-type [19] potentials where non-native contacts are completely ignored. The problem for this type of model is that the prediction for structures is based on a prior knowledge of the naive structures of protein. There are also reduced models that succeeded in folding helix bundles by using hydrogen bonds [20-23] and also succeeded in folding the beta hairpin by using specific dihedral potentials [24,25]. Even though both the helix and the beta sheet can be formed in this approach, they are folded separately using different potentials. In other words, bias potentials are often involved to achieve the folding. There are no known unbiased effective potentials that have been demonstrated to fold a structure with both of the helix and sheet structures on the computer.

In this thesis, we propose a reduced model that can fold both helix and sheet structures. The goal of this model is to find an unbiased effective potential that has included the effects of water and govern residues in the appropriate coarse-grained level. Because water molecules are integrated out, the resulting effective potential has to include the entropy effect and the dynamics due to water molecules. Therefore, in addition to contributions originated from electrostatic potentials, the effective potential also contains contributions purely from entropy effect and may have a total different form from microscopic considerations. An example is the so-called hydrophobic interaction that appears only when water molecules are integrated out and does not show up explicitly in microscopic consideration. The difficulty in this approach lies in the fact that there is no systematic way for constructing such potentials from the level of atoms. Experimentally, there is also not enough information that allows one to pin down the effective potential. Nevertheless, there has already accumulated useful analyses in the past that could shed light on what the potential are. For instance, the analysis on the Miyazawa-Jernigan (MJ) potential [26] shows that dipole-dipole interaction is still the dominating interaction between residues [27]. The effect of water molecules in the long-distance interaction may thus be captured by simple screening effect. Therefore, it suggests that by appropriate renormalization of parameters, microscopic forms for various interactions can be retained. Obviously, there is no reason why such renormalization can be carried down to the short distance. In the short-distance, the individuality of each water molecule begins to matter. The form of the potential may change violently. Indeed, we shall see that some potential does need a major modification when coming into short distance. The essence of our approach is to find the most unbiased potential that incorporates all known analyses on the effective potential. If the contributions have microscopic origin from electrostatic interactions, we apply the renormalized form with a global undetermined scale to be determined by experiments. If the contributions result from entropy effects, we analytically continue it to large distance with a global undetermined scale too. Using a few proteins with known structures for calibration, these undetermined scales can be fixed. By doing so, we are able to construct an unbiased potential that can fold both the helix and the beta sheets successfully. Our results indicate that the hydrogen bonding, the hydrophobicity, and the dipole-dipole interactions are three most important ingredients for determining the second structure. Note that the dipole-dipole interaction is the main ingredient in our effective potential that is different from others. Specifically, we find that the hydrogen bonding is essentially for the formation of the alpha helix. While in order to fold the beta sheet without bias, the dipole-dipole interaction is indispensable. As the potential we constructed is unbiased and is also allowed for further unbiased adjustment, it thus provides a reasonable and good start for solving the protein-folding problem.

This thesis is organized in the following way: In Chapter 2, we illustrate both biological and physical background knowledge for the protein structures. In Chapter 3, a detailed description of the proposed Reduced Protein Folding Model (RPFM) will be presented. The RPFM is a reduced off-lattice model in which peptide chains are represented by backbone structures with explicit structure and simplified side chain units. The degrees of freedom for backbones are based on the Ramachandran angles $\phi$ and $\psi$. In order to reduce the complexity and the time for computation, there is no internal degree of freedom for side-chains. Since water molecules are not included explicitly in this model, their effects are incorporated into several effective potentials. The first important interaction is the hydrogen bond interaction. This potential has been considered as the major interaction responsible for stabilizing secondary structures [20-23]. The second one is the hydrophobic interaction which is a mesoscopic potential induced by collective motion of water molecules, and is considered responsible for the compact globule formation of peptides [23]. In addition, two new interactions are introduced in our works: *dipole-dipole interactions* and *local hydrophobic interactions*. The forms for these two potentials are proposed by explicitly considering detailed structures of proteins. Their relevancy to the folding problem will be analyzed in great details. In Chapter 4, results of simulation for artificial peptides, *de novo* designed peptides and real protein peptides are demonstrated. The structures in these examples can be classified into several different types: one alpha helix, one beta sheet, one alpha helix and one beta sheet, and alpha helix bundles. Each of them has very different statistical properties and folding behavior. The energy landscapes and the Monte Carlo evolution are analyzed in this chapter. In addition, the effects of two new interactions, *dipole-dipole interactions* and *local hydrophobic interactions*, are discussed, too. In Chapter 5, we conclude with discussions on possible generalization. The Appendix collects the visual forms and the contact maps for all simulated native states with a comparison to the states that are either adapted from experiments or generally are believed to be the correct native states.

# Chapter 2

# Essentials of Proteins Structures and Numerical Methods

In this Chapter, we brief the essentials of proteins structures and numerical methods that are relevant in our later discussions and presentation of our main results. Here the essential of our major numerical method, the Monte Carlo method, will be also presented in Sec. 2.2.

## §2.1　Building Blocks and Level of Protein Structures

The building blocks of proteins are amino acids. In nature, only twenty different kinds of amino acids are incorporated into proteins. The complete structures of these amino acids are shown in the Fig. 2.1. Amino acids consist of two major parts: one is the backbone (gray areas shown in Fig. 2.1) and the other is the side-chain (orange areas in Fig. 2.1). The backbones in amino acids are all the same and the difference between different amino acids lies in the side-chain that is often denoted by R. Due to the apparent differences in the side-chains, 20 amino acids have different properties and thus it allows huge of number of ways for constructing proteins.

　　Amino acids can join together to form a linear chain via the formation of *peptide bond*, as demonstrated in Fig. 2.2. The peptide bond can be considered a resonance hybrid of the following forms:

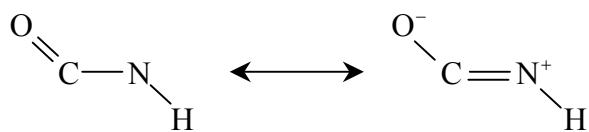

In other words, the peptide bond has partial double bond character. Hence, the six atoms of the peptide bond group are always coplanar and form the so-called *amide plane*, as shown in Fig. 2.2. When an amino acid is added to the chain from either end, a water molecule must be removed. The portion of each amino acid remaining in the chain is called an amino acid *residue*. As this reaction continues, the chain gets longer. This chain sometimes is called *polypeptide*, i.e. the protein itself and the sequence of amino acids that form the chain is the *primary structure* of protein.

After the peptide chain is formed, they start to form the local structures. The precise mechanism behind this formation is one of the main concerns in protein folding problem. Nevertheless, as early as 1953, it was already speculated by Pauling that the formation is done via the hydrogen bonds forming between backbones during the folding. These local structures are later referred as the so-called *secondary structure* of proteins and can be classified into several types: alpha helixes, beta sheets, and random coils. (See Fig. 2.3, the detailed graphic illustration of alpha helixes and beta sheets will be postponed to Chapter 4.) A standard alpha helix has 3.6 residues per turn and the hydrogen bond is formed between the CO group of the *i*th residue and the NH group of the (*i+4*)th residue, while for the beta sheets, they are composed by the so-called beta strands arranged in either parallel and anti-parallel fashion. The random coil is a protein or a segment of a protein that completely lacks secondary structure. Moreover, the denaturing of a protein reduces a protein entirely to random coil. Simple combinations of a few secondary structures are often seen in the proteins and hence are termed as motifs. A couple of frequent encountered motifs, including the hairpin β motif and the helix-turn-helix motif, are illustrated in Fig. 2.3.

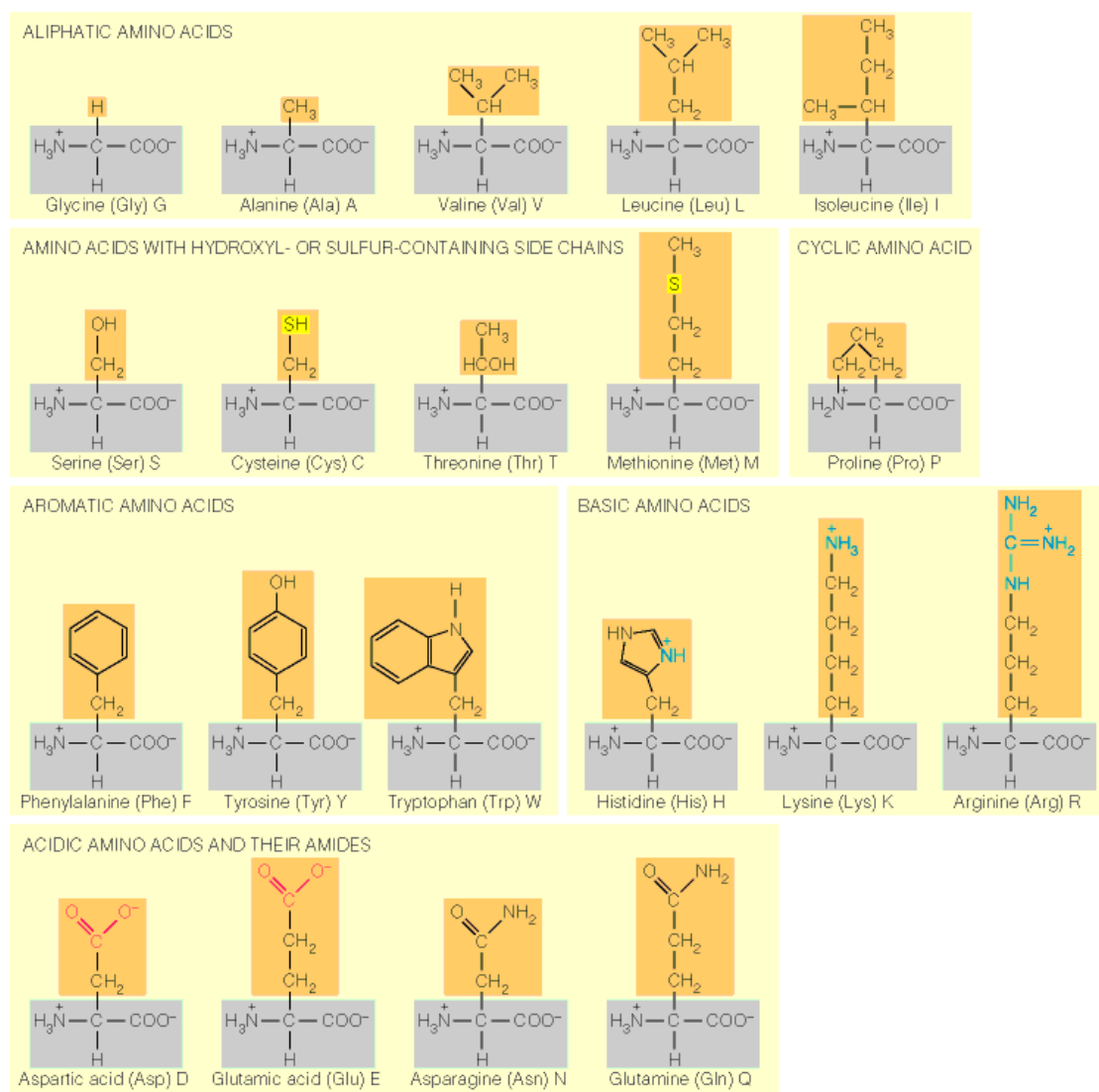

Fig. 2.1 The structures of 20 amino acids. The gray areas are the backbones and the orange areas are the side-chains. Adapted from [28].

The next level after the secondary of protein structures is the so-called *tertiary structure*. This is the overall three-dimensional conformation of protein. The main interaction that causes these structures is believed to be the hydrophobic collapse. After the tertiary structures are formed, two or more than two proteins sometimes can interact with each other. Hence, they can form the so-called *quaternary structure* of proteins. The hierarchy of these levels in proteins is illustrated in Fig. 2.4.

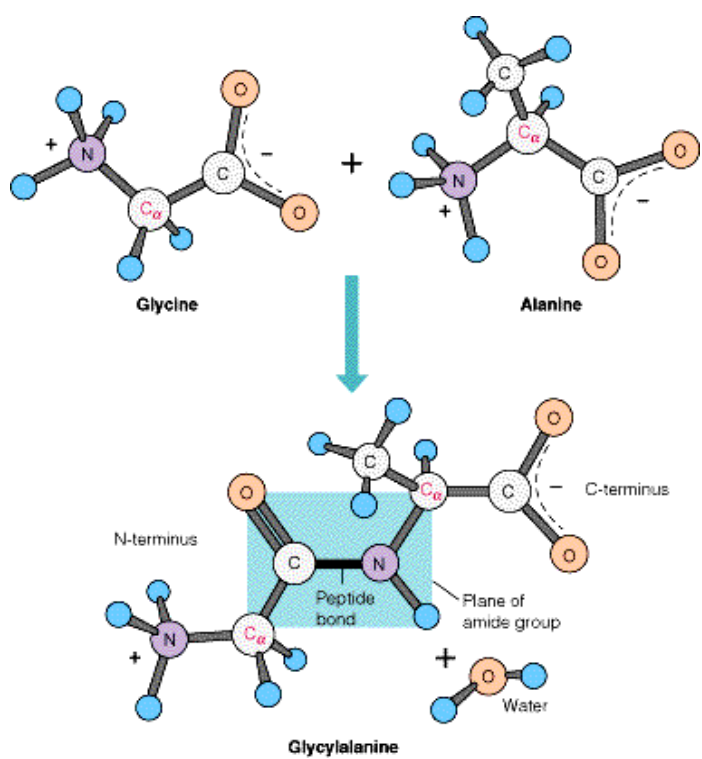

Fig. 2.2 The formation of a peptide bond. The cyan plane is the amide plane. Adapted from [28].

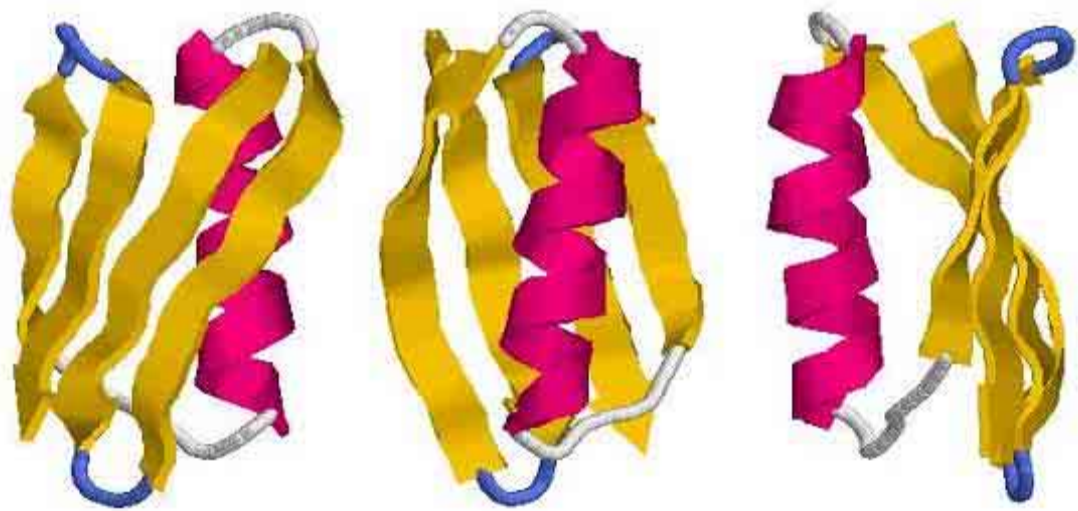

Fig. 2.3 Immunoglobulin Binding Protein (1PGA). This protein consists of alpha helix, parallel and anti-parallel beta sheets, and random coil structures. The pink part of this structure is a typical alpha helix. The yellow parts are parallel and anti-parallel beta sheets. The blue parts are the turn structures. And the white parts are the random coils.

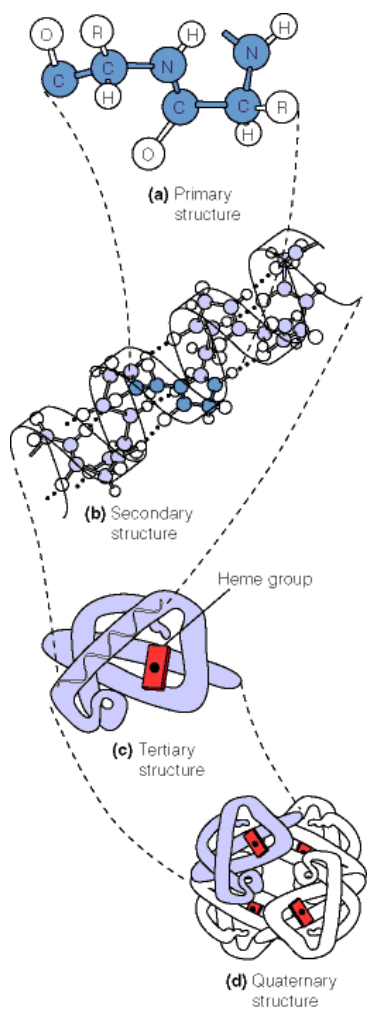

Fig. 2.4 The level of protein structures. Adapted from [28].

## §2.2 Monte Carlo Method

There are several numerical methods invoked for simulating protein folding. The principal simulation method in our work employs the Monte Carlo method, using the Metropolis algorithm. This algorithm was developed by Metropolis *et al.* in 1953 [29]. The basic idea of this algorithm is to generate the final state based on the previous one, using a transition probability that depends on the energy difference between the initial and final state. Thus the system can be considered as a stochastic system so that the time-dependent behavior of this system can be described by a master equation

$$\frac{\partial P_n(t)}{\partial t} = -\sum_{n \neq m} \left[ P_n(t) W_{n \to m} - P_m(t) W_{m \to n} \right], \tag{2.1}$$

where $P_n(t)$ is the probability of the system being in state *n* at time *t*, and $W_{n \to m}$ is the transition rate for $n \to m$. When the system is in equilibrium, $\partial P_n(t)/\partial t = 0$ and the two terms on the right-hand side of Eq. (2.1) must be equal. The resulting expression is known as the detailed balance condition

$$P_n(t) W_{n \to m} = P_m(t) W_{m \to n}. \tag{2.2}$$

The condition of detailed balance implies that, at equilibrium, the average probability of moves $n \to m$ is the same as the average probability of inverse moves $m \to n$. The probability of the *n*th state occurring in a canonical ensemble system is given by

$$P_n(t) = \frac{e^{-\frac{E_n}{kT}}}{Z}, \tag{2.3}$$

where *Z* is the partition function. $k$ is the Boltzmann constant. *T* is the temperature of the system. This probability is usually not exactly known because sometimes the complete enumeration of states is impossible, i.e. *Z* is usually unknown. However, one can avoid this situation by generating each new state directly from the preceding state. If the *n*th state is generated from the *m*th state, the relative probability is the ratio of the individual probabilities and the partition function *Z* cancels. As a result, only the energy difference between the two states is needed, hence

$$\frac{W_{n \to m}}{W_{m \to n}} = e^{-\frac{\Delta E}{kT}}, \quad (2.4)$$

where $\Delta E = E_n - E_m$. Any transition rate that satisfies detailed balance is acceptable. The Metropolis algorithm choose the following expression for the probability of acceptance

$$W_{n \to m} = \min\left\{1, e^{-\frac{\Delta E}{kT}}\right\}. \quad (2.5)$$

It is easy to prove that the Eq. (2.5) for $W_{n \to m}$ which was first proposed by Metropolis *et al.* satisfies the Eq. (2.4). [30] The Metropolis algorithm in our model is listed as follows:

*Step 1*: Choose an initial conformation that is generated randomly. Calculate its energy.

*Step 2*: Generate a new conformation by making a small change in the Ramachandran angles $\phi$ or $\psi$. Calculate the energy difference $\Delta E$ between two conformations.

*Step 3*: If $\Delta E < 0$, then the resulting conformation is accepted.
If $\Delta E > 0$, generate a random number $R$ between 0 and 1, i.e. $0 < R < 1$.
If $e^{-\frac{\Delta E}{kT}} > R$, then the resulting conformation is accepted.
If $e^{-\frac{\Delta E}{kT}} < R$, then the resulting conformation is refused.

*Step 4*: Go to *Step 2* until the native state is reached.

# Chapter 3

# Reduced Protein Folding Model

In this chapter, a detailed description of the proposed Reduced Protein Folding Model (RPFM) will be presented. The RPFM is a reduced off-lattice model in which peptide chains are represented by backbones with explicit structures and simplified side-chain units. The degrees of freedom for backbones are based on the Ramachandran angles $\phi$ and $\psi$. In order to reduce the complexity and the time for computation, there is no internal degree of freedom for side-chains. Meanwhile, water molecules are not included explicitly in this model, but their effects are incorporated into several effective potentials. The first important interaction is the hydrogen bond interaction. This potential has been considered as the major interaction responsible for stabilizing secondary structures [20-23]. The second one is the hydrophobic interaction. The hydrophobic effect is a mesoscopic force induced by collective motion of water molecules, and it is considered responsible for the compact globule formation of peptides. The so-called desolvation model [31-34] is a typical model that represents this effect approximately. In addition, two new interactions are introduced in our works, the RPFM: *dipole-dipole interactions* and *local hydrophobic interactions*. As we shall show, the forms for these two potentials are proposed by explicitly considering detailed structures of proteins. The existence of these two is the main different feature between the current model and others. With these two interactions, our model, RPFM, can simultaneously fold both helix and sheet structures.

In Sec. 3.1, we illustrate the chain representation and degrees of freedom. In Sec. 3.2, quite detailed descriptions of interaction potentials are given. The potentials in RPFM consist of three main parts: steric interactions, directional interactions, and hydrophobic interactions. All of these interactions have both global and local forms. In Sec. 3.3, conformation parameters are described briefly. In addition to the radius of gyration ($R_G$) and native contact number ($Q$), a new parameter, water accessible volume (*WAV*), is introduced for depicting energy landscape.

# §3.1 Coarse-Grained Representation of the Protein Molecules

## §3.1.1 Backbone Units

Peptide chains in our model are represented by two components: backbone units and side-chains. First, backbone units are represented by spheres, as shown in Fig. 3.1. The diameter of these spheres is defined by the distance between two nearest successive $C^\alpha$-atoms. This distance is about 3.7842Å on the average based on analysis of Protein Data Bank (PDB) data. Each unit contains five atoms (C, O, N, H and $C^\alpha$-atoms) with fixed bond lengths and bond angles. In addition, redundant atoms are added for consistency of backbone units as indicated outside the dash straight line in Fig. 3.1. We thus require the number of backbone be more than the number of amino acid residue by one to include these redundant atoms in the N- and C-terminal.

As mentioned earlier, the peptide bond (C─N) have partial double-bond character. Therefore, the atoms in backbone unit are coplanar and form the so-called amide plane. Consider an amide plane that lies on *x-y* plane. Let the carbon atom on the peptide bond is at the origin of coordinate and peptide bond is in the +*x* direction. The relative coordinates of these atoms could be determined from fixed bond lengths and bond angles, as listed in Table 3.1 and shown in Fig. 3.2. These relative coordinates are used to define hydrogen bonds and the locations of dipoles in the corresponding interactions. One of the main reasons for using these coordinates is to provide an important fixed reference so that the accumulation of round off error when doing Monte Carlo moves would not happened. Due to Monte Carlo move, if one naively uses fixed bounds and rotation angles, round off error would accumulate and eventually affect the rotational motion of the peptide chains.

The main degrees of freedom for backbone unit are based on the Ramachandran angles $\phi_i$ and $\psi_i$, as illustrated in Fig 3.3. $\phi_i$ is the angle defined by atoms $C_{i-1}─N_i─C^\alpha_i─C_i$ and $\psi_i$ is the angle defined by atoms $N_i─C^\alpha_i─C_i─N_{i+1}$ in each unit (except for the N- and C-terminal). $\omega_i$ is the angle around the peptide bond. It is restricted to 180° as the result of the partial double-bond character of peptide bond.

Why do we restrict $\omega$ is restricted to be 180°? Typically, $\omega = 0°$ corresponds to *cis* conformation of amide plane and $\omega = 180°$ corresponds to *trans* one, as shown in Fig. 3.4. For *cis* conformation, the distance between two $C^\alpha$-atoms is around 2.8Å. This is less than the minimum distance allowed for the C…C pair (the minimum van der Waals radius of two carbon atoms is 3.0Å, see Table 3.4). As for *trans* one, this distance is about 3.8Å (i.e. the diameter of our backbone unit). $C^\alpha$-atoms of the sequence-neighboring amino acids are far apart in space due to the rigid *trans* form of the peptide bond, and this provides an opportunity for these neighboring residues to change their conformations almost independently of each other. Hence, $\omega = 180°$ is allowed and $\omega = 0°$ is disallowed for all amino acid residues except proline. We will discuss proteins with prolines later. Another consideration is that the potential barriers between *trans* and *cis* conformations are high due to the double-bond character of peptide bond. As a result, the typical range rotation around the peptide bond due to thermal fluctuation is less than 10° [37]. Overall speaking, in order to reduce the computing time, we fix $\omega$ to be 180° in our model so that the final conformation corresponds to the *trans* one.

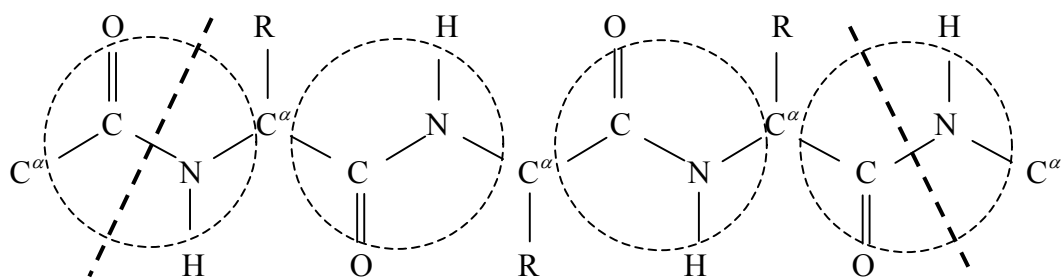

Fig. 3.1 Backbone units are represented by spheres with diameter 3.7842Å. Each unit contains five atoms: C, O, N, H and $C^\alpha$-atoms. R represents side-chain which is attached to the $C^\alpha$-atom in a rigid way. Redundant atoms are added for consistency of backbone units. These redundancies are outside the dash straight lines in the two ends.

Table 3.1 Relative coordinates, bond lengths and bond angles for backbone unit [35].

| Atom coordinates | | Bond lengths, Å | | Bond angles, ° | |
|---|---|---|---|---|---|
| $C_i$ | ( 0.000, 0.000) | $C^\alpha C$ | 1.510 | $C^\alpha CN$ | 116.0 |
| $O_i$ | (−0.684, 1.034) | CN | 1.325 | $C^\alpha CO$ | 120.5 |
| $C^\alpha_i$ | (−0.662, −1.357) | CO | 1.240 | OCN | 123.5 |
| $N_{i+1}$ | ( 1.325, 0.000) | NH | 1.020 | CNH | 119.5 |
| $H_{i+1}$ | ( 1.827, −0.888) | $NC^\alpha$ | 1.455 | $CNC^\alpha$ | 122.0 |
| $C^\alpha_{i+1}$ | ( 2.096, 1.234) | $C^\alpha C^\beta$ | 1.540 | $HNC^\alpha$ | 118.5 |
| $C_{i+1}$ | ( 3.578, 0.946) | | | $NC^\alpha C$ | 111.0 |
| | | | | $NC^\alpha C^\beta$ | 110.0 |
| | | | | $CC^\alpha C^\beta$ | 110.0 |

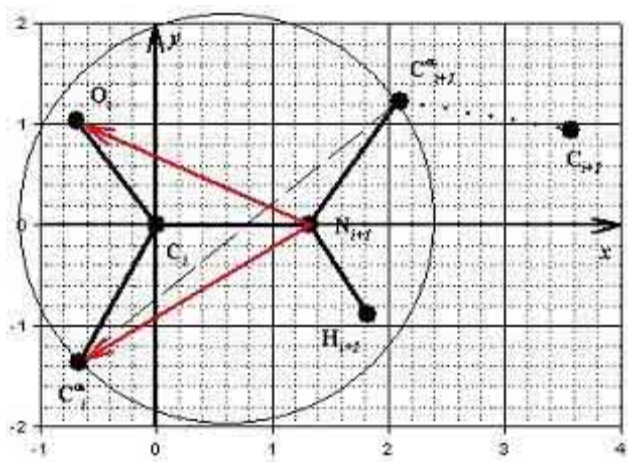

Fig. 3.2 Figure of relative coordinates for each atom in one backbone unit. Circle represents the backbone unit and dash line is its diameter. Sub-indices indicate that atoms are from the *i*th or (*i+1*)th amino acid residues. The two vectors shown in the figure are used for defining the normal vector of the amide plane. The relevant coordinate data are listed in Table 3.1.

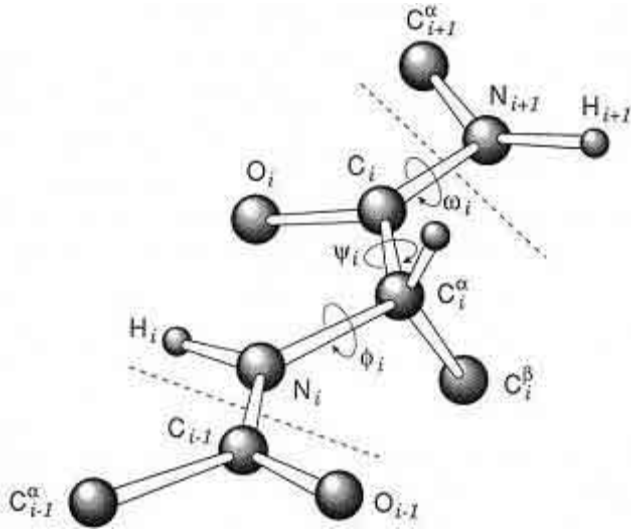

Fig. 3.3 Main degrees of freedom for a backbone unit are based on the Ramachandran angles $\phi_i$ and $\psi_i$. $\omega_i$ is restricted to 180° for *trans* conformation of amide plane. Adapted from [36].

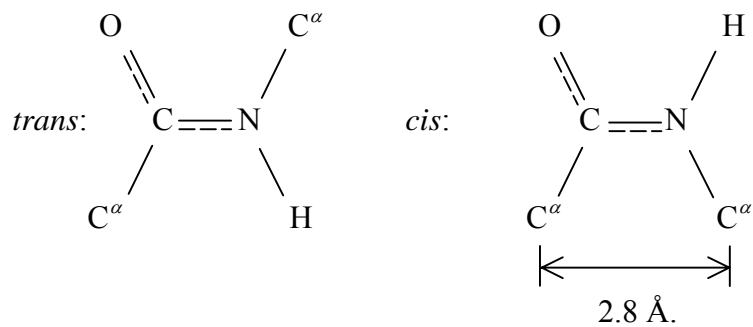

Fig. 3.4 *Trans* and *cis* conformations of amide plane correspond to $\omega = 180°$ and $\omega = 0°$, respectively. *Trans* is allowed and *cis* is disallowed for all amino acids because the distance of two neighboring $C^\alpha$-atoms is less than the minimum van der Waals radius of two carbon atoms in *cis* conformation.

### §3.1.2 Side-chain Units

Side-chains, in our coarse-grained approach, are simply represented as spheres with specific effective radii. Their locations are just at $C^\beta$-atoms that attached to $C^\alpha$-atoms in a rigid way, as shown in Fig. 3.3. Average speaking, the distance from $C^\alpha$ to $C^\beta$ is found to be 1.540Å approximately, using the PDB data. In addition, coordinates of $C^\beta_i$ can be also inferred from knowing the coordinates of $N_i$, $C^\alpha_i$, and $C_i$ atoms. In order to reducing the complexity of simulation we do not use any internal degree of freedom for side-chain units in RPFM.

Different amino acid residues have different effective global and local radii in the model. When residues are far apart, one expects that internal structures of residues are not important. This reflects in that each residue can be described by its own global radius. This effective radius can be determined by the *Langevin's equation* experimentally [40]. However, there is also an empirical formula to the radius:

$$R = \left( \sqrt[3]{\frac{3}{4\pi}(M.W. \times 1.3 \times \delta)} \right) \times \gamma, \tag{3.1}$$

where M.W. is the molecular weight of amino acid residue (in daltons). Notice that the molecular weight calculated here does not contain the atoms in backbone. $\delta$ is a dimensionless factor for correcting volumes. If amino acid residue contains many aliphatic ($-CH_2-$, $-CH_3$) groups, then $\delta = 1.05$. If residue contains many polar (O, N) atoms, then $\delta = 0.95$ [37]. $\gamma$ is another dimensionless factor for correcting. Consider three amino acids: VAL, THR, and ILE. All of them have two carbons, $C^\gamma$ and $C^\delta$, on their $C^\beta$-atom, others have only one, $C^\gamma$-atom (see Fig. 2.1). Due to the steric exclusion of these two carbons, their side chain radii need to be enlarged. The radius enlargement is around 10% from our simulation results, i.e. $\gamma = 1.10$. For those amino acids that contain benzene ring, since the benzene rings could be stacked in a more compact way, these aromatic amino acids could be much closer in distance. Therefore, their radius reduction is about 85% from simulation results, i.e. $\gamma = 0.85$. These amino acids are: HIS, PHE, TYR, and TRP. All effective global radii are tabulate in the Table 3.2.

When the amino acid residues come close, the effective global radii cease to work. The reason is simple because each residue now begins to see the internal structure of other residues, and residues may stack in different way. Therefore there is no a prior reason the potential that residues interact in large distance and small distance can be fit into single effective radius. In fact, the global radii calculated using Eq. (3.1) may also be either too large or too small for local interactions. For this above reason, to account for local interactions, we use different radius to describe interaction between $i$th and ($i+1$)th amino acid residues. Roughly speaking, the local radius of residue can be classified into two: $R_{big}$ and $R_{small}$. This comes from the consideration of connection of carbon atoms in residues. For residues, with serial carbon connection, they are smaller and hence we assign $R_{small}$. For others, we assign $R_{big}$. As a result, in our model, we assign VAL, THR, ILE, and TYR $R_{big} = 2.87 \text{ Å}$ and assign others $R_{small} = 2.60 \text{ Å}$ for their effective local radius.

According to textbooks [38] and our simulation experience, amino acids may be divided into five classes for side-chain properties in RPFM: *hydrophobic* (H), *polar* (P), *neutral* (N), *positive charged* ($+$), and *negative charged* ($-$). These properties will be used for constructing local hydrophobic interactions. The first class comprises those with strictly hydrophobic side-chains: ALA, VAL, CYS, LEU, ILE, MET, PHE, and TRP. The second class comprises those with polar side-chains: SER, THR, ASN, GLN, HIS, and TYR. Additionally, two amino acids glycine (GLY) and proline (PRO) are special residues. Glycine, which has only one hydrogen atom as a side-chain, possesses special properties. Proline, which side-chain connects back to its backbone, has a specific backbone conformation. Thus, these two amino acids are considered to form the neutral class. The last four amino acids are charged residues with positive or negative charge on their side-chains. LYS, ARG are in the positive class and ASP, GLU are in the negative, respectively. Side-chain properties of each amino acid are classified accordingly in the Table 3.2.

Among the above classification, two amino acids in the neutral class need to be considered separately. Even though the side-chain of glycine contains only one hydrogen atom, its side-chain center is still at the location of $C^{\beta}$-atom. The only difference is that one need a smallest radius 0.6770Å both globally and locally for consistency of programming. In the case of proline, it has a specific backbone

Table 3.2 Molecular weight (M.W., in daltons), volume correction factor ($\delta$, dimensionless), radius correction factor ($\gamma$, dimensionless), effective global radius (in Å), effective local radius ("S" and "B"), and side-chain (S.C.) properties for each amino acid (A.A.), sorted by molecular weight. Notice that the molecular weight calculated here does not contain the atoms in backbone. The abbreviations in effective local radius, "S" and "B" mean $R_{small} = 2.60\,\text{Å}$ and $R_{big} = 2.87\,\text{Å}$, respectively. In the column of S.C. properties, "H", "P", "N", "+", and "−" represent *hydrophobic*, *polar*, *neutral*, *positive charged*, and *negative charged* side-chain properties, respectively. The radius of backbone (B.B.) unit is on the bottom of table. This radius is just defined by the half distance between two nearest successive $C^\alpha$-atoms.

|      |      | Correction factor | | Radius, Å | | S.C. |
| --- | --- | --- | --- | --- | --- | --- |
| A.A. | M.W. | $\delta$ | $\gamma$ | Global | Local | properties |
| GLY | 1   | 1.0000 | 1.00 | 0.6770 | S | N |
| ALA | 15  | 1.0000 | 1.00 | 1.6697 | S | H |
| SER | 31  | 0.9500 | 1.00 | 2.0908 | S | P |
| PRO | 41  | 1.0500 | 1.00 | 2.3729 | S | N |
| VAL | 43  | 1.0500 | 1.10 | 2.6519 | B | H |
| THR | 45  | 0.9975 | 1.10 | 2.6468 | B | P |
| CYS | 47  | 1.0000 | 1.00 | 2.4433 | S | H |
| LEU | 57  | 1.0500 | 1.00 | 2.6483 | S | H |
| ILE | 57  | 1.0500 | 1.10 | 2.9132 | B | H |
| ASN | 58  | 0.9500 | 1.00 | 2.5763 | S | P |
| ASP | 58  | 0.9500 | 1.00 | 2.5763 | S | − |
| GLN | 72  | 0.9975 | 1.00 | 2.8143 | S | P |
| GLU | 72  | 0.9975 | 1.00 | 2.8143 | S | − |
| LYS | 73  | 0.9975 | 1.00 | 2.8272 | S | + |
| MET | 75  | 1.0500 | 1.00 | 2.9020 | S | H |
| HIS | 81  | 0.9975 | 0.85 | 2.4879 | S | P |
| PHE | 91  | 1.0500 | 0.85 | 2.6310 | S | H |
| ARG | 101 | 0.9975 | 1.00 | 3.1504 | S | + |
| TYR | 107 | 0.9975 | 0.85 | 2.7298 | B | H |
| TRP | 130 | 0.9975 | 0.85 | 2.9129 | S | H |
| B.B. | 56  |        |      | 1.8921 |   |   |

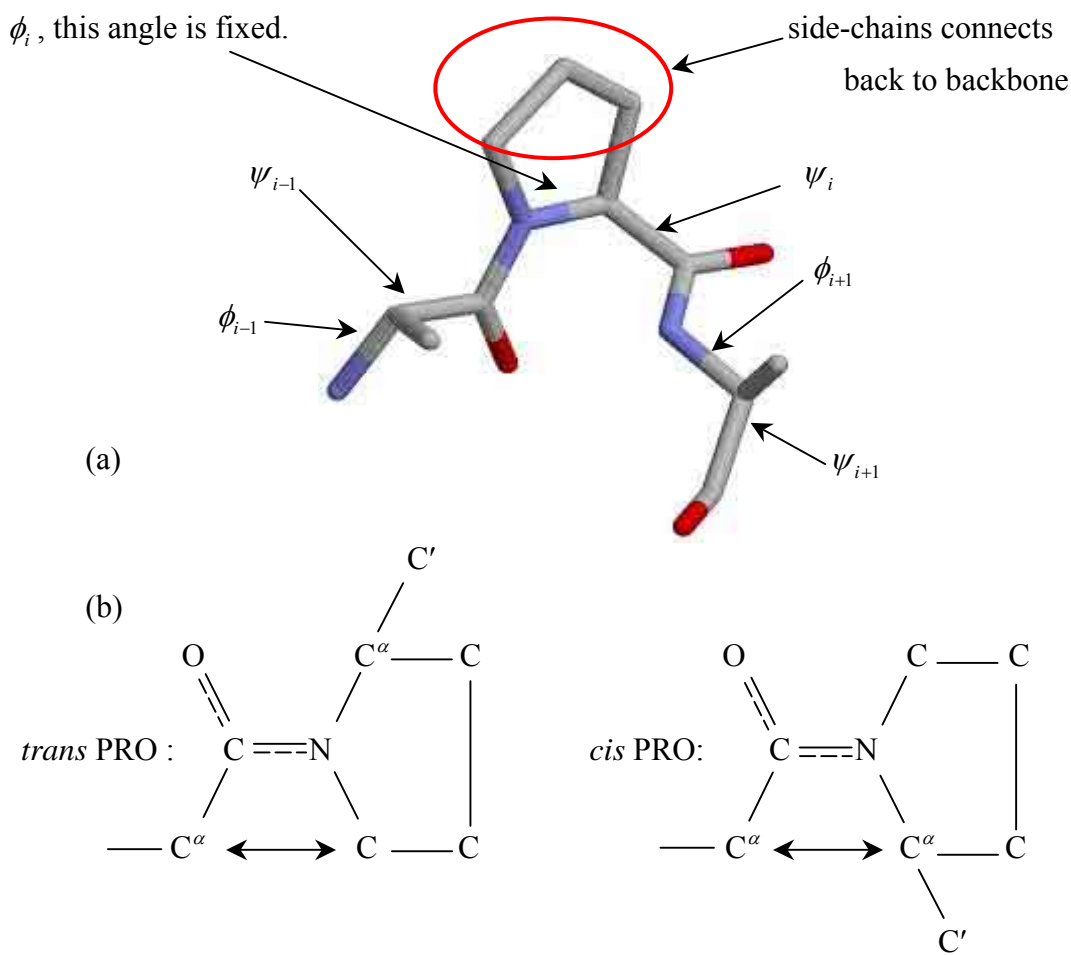

(a)

(b)

trans PRO :   
$$\begin{array}{c} O \\ \| \\ C \end{array} \begin{array}{c} C^\alpha — C \\ / \\ === N \end{array}$$
$$—C^\alpha \longleftrightarrow C — C$$

cis PRO:
$$\begin{array}{c} O \\ \| \\ C \end{array} \begin{array}{c} C — C \\ / \\ === N \end{array}$$
$$—C^\alpha \longleftrightarrow C^\alpha — C$$
$$\begin{array}{c} \\ C' \end{array}$$

Fig. 3.5 (a) The backbone structure of PRO. $\phi_i$ is fixed to $-60°$ owing to its side-chain connecting back to backbone, as shown in the circle. (b) Proline in *trans* and *cis* conformations. Both of them have carbon atom in the neighbor. Hence, *trans* conformation has no advantage as compared with the *cis* one.

conformation owing to its side-chain connecting back to backbone. Therefore, there is no degree of freedom for $\phi_i$ when proline is appearing in side-chains, as illustrated in Fig. 3.5 (a). In some textbooks [36-38], $\phi_i$ is given in the range of $-65° \sim -70°$. Statistical analysis from PDB on $\phi_i$ gives in the range of $-55° \sim -75°$. Therefore, $\phi_i$ is almost fixed in proline. Furthermore, according to the $sp^3$-hybridization of carbon, the potential of dihedral angle of carbon has three minimums within $360°$: $60°$, $180°$, and $300°$ (i.e. $-60°$), respectively. Hence, $\phi_i$ is set to a fixed value $-60°$ in our model. Finally, there are *trans* and *cis* conformations for amide plane as mentioned previously. Proline has *trans* and *cis* conformations, too. But its *trans* conformation has only minor advantage as compared with the *cis* one. This is because both of these conformations have carbon atoms in the neighbor (see Fig. 3.5 (b)). In both globular and unfolded peptides, there are about 90% of *trans* and 10% of *cis* proline [37]. Therefore, in order to reducing computation again, the *trans* proline is adopted in our model.

# §3.2 Interaction Potentials

After defining the coarse-grained units in our model, we now turn to describe the interaction among these units. Our potential consists of three major parts: steric interactions, directional interactions, and hydrophobic interactions,

$$V_{total} = V_{steric} + V_{directional} + V_{hydrophobic}. \qquad (3.2)$$

Each of them has both global and local terms. The local potential terms mean interacting between nearest successive units, i.e. *i*th and (*i+1*)th backbone or side-chain units, while the global ones are not restricted to nearest neighbor interactions. The others belong to global terms. We will interpret them one by one in the following.

First we consider the steric interactions, $V_{steric}$, which consists of three terms,

$$V_{steric} = V_{SA} + V_{dihedral} + V_{avoiding}. \qquad (3.3)$$

The self-avoidance (*SA*) term $V_{SA}$, which is a global interaction, is given by a hard-sphere potential of the form

$$V_{SA} = \varepsilon_{SA} \times \sum_{i,j} \left( \frac{\sigma_i + \sigma_j}{r_{ij}} \right)^{12}, \qquad (3.4)$$

where $\sigma_i$ and $\sigma_j$ are the global radii of backbone or side-chain units in the Table 3.2. $r_{ij}$ is the distance between units. The sum runs over all non-successive unit pairs, such as backbone vs. backbone, side-chain vs. side-chain, or backbone vs. side-chain. $\varepsilon_{SA}$ is the energy reference; this would be discussed later.

The dihedral term $V_{dihedral}$, which is a local interaction, assigns real dihedral energy according to the configuration of dihedral angles. In our model, this potential is not the standard form with the threefold symmetry [39] (One of $sp^3$ angle is excluded because it is out of range.) It is designed by the relative orientation of two successive amide planes, in other words, the relative orientation of two successive backbone units in RPFM. Any two successive amide planes with non-neutral side-chain between them can have this interaction. Statistical analysis of this relative orientation in some peptides shows that they have two distributions. One distribution prefers helix structures and the other prefers sheets, we shall term them as Dihedral A (*DA*) and Dihedral B (*DB*), respectively. When the orientation of *i*th amide plane is given, the (*i+1*)th amide plane either Dihedral A or Dihedral B. Now, it becomes an analytic geometry problem: given a plane, how many parameters are necessary for determining another plane for a specific orientation? Consider two successive amide planes, there are twenty-five distances connecting atoms in different plane excluding the joint $C^\alpha$-atoms. This orientation problem could be solved by picking any three distances from non-concurrent atoms in each plane. For Dihedral A, namely helixes like, distances of $O_i \ldots O_{i+1}$, $H_i \ldots H_{i+1}$, and $C^\alpha_i \ldots C^\alpha_{i+1}$ are chose. For Dihedral B, i.e. sheets like, distances of $H_i \ldots O_{i+1}$, $O_i \ldots H_{i+1}$, and $C_i \ldots N_{i+1}$ are used. These statistical values are listed in the Table 3.3. In this table, $r_{ave}$ means the average distance of two atoms when orientation of two successive amide planes is in either Dihedral A or Dihedral B. $r_{max}$ and $r_{min}$ are the maximum and minimum distances in all possible orientation of two amide planes in PDB, respectively. $r_{range}$ is the maximum value of $|r_{ave} - r_{max}|$ or $|r_{ave} - r_{min}|$.

The chiral problem that there are only L-type residues in peptides is also solved in this potential. If we define the normal vector of the (*i+1*)th amide plan (see Fig. 3.2) as

$$\mathbf{n}_{i+1} = \mathbf{r}_{N_{i+1}C^\alpha_i} \times \mathbf{r}_{N_{i+1}O_i} \tag{3.5}$$

and consider the relative orientation of nitrogen atom and normal vector in two successive amide planes, there are four possibilities of their orientations as illustrated in Fig. 3.6. When $N_{i+1}$ atom is in the positive side of the *i*th amide plane, either $V_{DihedralA}$ or $V_{DihedralB}$ would be selected. Then, by the same token, when $N_i$ atom is in the positive side of the (*i+1*)th amide plane, we take $V_{DihedralA}$, otherwise we take $V_{DihedralB}$. Therefore, the dihedral term $V_{dihedral}$ has the form

$$V_{dihedral} = \begin{cases} 0 & \text{when the orientation is (a) or (b) in Fig. 3.6} \\ V_{DihedralA} & \text{when the orientation is (c) in Fig. 3.6} \\ V_{DihedralB} & \text{when the orientation is (d) in Fig. 3.6} \end{cases} \quad (3.6)$$

with

$$V_{DihedralX} = \begin{cases} \varepsilon_{DX} \times \sum_{\substack{3\,pairs \\ i,i+1}} \left[ \left( \dfrac{r_{i,i+1} - r_{ave}}{r_{range}} \right)^2 - 1 \right] & \text{when } |r_{i,i+1} - r_{ave}| \leq r_{range} \\ 0 & \text{when } |r_{i,i+1} - r_{ave}| > r_{range} \end{cases}, \quad (3.7)$$

where X represents "A" or "B". When $V_{DihedralA}$ is selected, three pairs that the sum runs over are in the top three rows of Table 3.3. When $V_{DihedralB}$ is taken, three pairs are in the bottom three. $r_{i,i+1}$ is the distance between these pairs. The summation runs over all successive backbone units with non-neutral side-chain between them. $\varepsilon_{DX}$ is the constant dihedral strength. It should be stressed that $\varepsilon_{DA}$ and $\varepsilon_{DB}$ could be different when considering all interactions.

The last term in the steric interactions, $V_{avoiding}$, is used for avoiding the unreasonable physical situations, it consists of four terms

$$V_{avoiding} = V_{\phi_i \psi_i} + V_{BB_i BB_{i+1}} + V_{H_i H_j} + V_{O_i O_j}. \quad (3.8)$$

The first two terms, $V_{\phi_i \psi_i}$ and $V_{BB_i BB_{i+1}}$, are local interactions. $V_{\phi_i \psi_i}$ eliminates the conformation of $\phi_i = 0°$ and $\psi_i = 0°$. This conformation is not allowed for any peptides because of the steric clash between the carbonyl oxygen and amino proton, see Fig. 3.7. Clearly, $V_{\phi_i \psi_i}$ is imposed between the oxygen atom in the *i*th amide plane and the hydrogen atom in the (*i+1*)th amide plane. The interaction $V_{BB_i BB_{i+1}}$ is

Table 3.3 Three distance pairs used for $V_{DihedralA}$ and $V_{DihedralB}$. The top three rows are used for $V_{DihedralA}$ and the bottom three are for $V_{DihedralB}$. Average, maximum, and minimum of $r$ (in Å) are statistical data from these atom pairs in two successive amide planes in PDB. $r_{range} = \max\{|r_{ave} - r_{max}|, |r_{ave} - r_{min}|\}$.

| $V_{DihedralA}$ | $r_{ave}$ | $r_{max}$ | $r_{min}$ | $r_{range}$ |
|---|---|---|---|---|
| $O_i \ldots O_{i+1}$ | 3.2793 | 5.1864 | 2.3370 | 1.9071 |
| $H_i \ldots H_{i+1}$ | 2.7312 | 4.8189 | 1.8685 | 2.0877 |
| $C^\alpha_i \ldots C^\alpha_{i+1}$ | 5.3595 | 7.2586 | 4.8333 | 1.8991 |
| $V_{DihedralB}$ | $r_{ave}$ | $r_{max}$ | $r_{min}$ | $r_{range}$ |
| $H_i \ldots O_{i+1}$ | 2.5419 | 4.5723 | 2.0041 | 2.0304 |
| $O_i \ldots H_{i+1}$ | 4.3833 | 5.2608 | 1.9090 | 2.4743 |
| $C_i \ldots N_{i+1}$ | 4.4209 | 4.7863 | 2.6585 | 1.7624 |

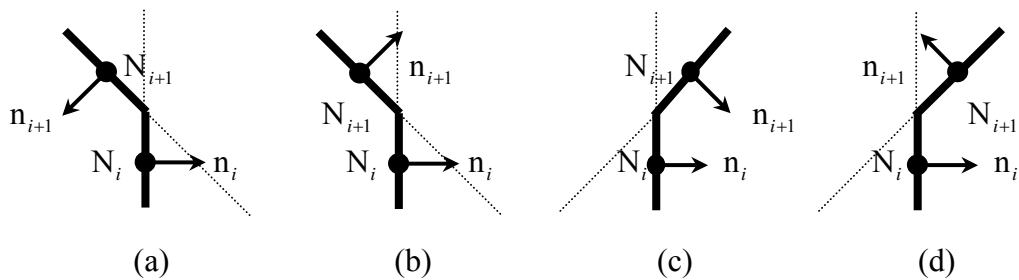

(a)　　　　(b)　　　　(c)　　　　(d)

Fig. 3.6 (a) and (b) represent the relative orientation of two neighboring amide planes in the D-type residues. There is no dihedral energy when two amide planes are in these two orientations. (c) and (d) represent the helix and sheet preferring orientations, respectively. When two amide planes are in (c) orientation, $V_{DihedralA}$ would be selected. When in (d) orientation, $V_{DihedralB}$ would be taken.

just the self-avoidance term between two nearest successive backbone units, which is not taken into account in the $V_{SA}$. The last two terms, $V_{H_iH_j}$ and $V_{O_iO_j}$, are global interactions among any $H_i \ldots H_j$ pairs or $O_i \ldots O_j$ pairs, respectively. Owing to our model, the hydrogen atoms or oxygen atoms in different backbone units might get too closer in distance. Hence, $V_{H_iH_j}$ and $V_{O_iO_j}$ are used for excluding this situation.

All of these four terms have the same form shown in as Eq. (3.4) with slightly changes. In others words, these potentials are just the self-avoiding interactions in the atomic scales. Their minimum distances allowed are the sum of their van der Waals radii (The typical minimum van der Waals radii of each atom are listed in Table 3.4.) Hence, $V_{avoiding}$ is given by

$$V_{XY} = \varepsilon_{avoiding} \times \sum_{X,Y} \left( \frac{\sigma_X + \sigma_Y}{r_{XY}} \right)^{12}, \quad (3.9)$$

where $\varepsilon_{avoiding}$ is a scale, much larger than any other energies in our model. The van Waals radii used in different $XY$ have different meanings. When $XY = \phi_i\psi_i$, $\sigma_X$ and $\sigma_Y$ are the radius of $O_i$ and $H_{i+1}$, respectively. When $XY = BB_iBB_{i+1}$, $\sigma_X + \sigma_Y = 2.5$ Å. (Note that this distance is small than 3.7842Å, i.e. the diameter of backbone unit.). When $XY = H_iH_j$, $\sigma_X$ and $\sigma_Y$ are the radius of hydrogen atom. When $XY = O_iO_j$, $\sigma_X$ and $\sigma_Y$ are the radius of oxygen atom. $r_{XY}$ is the distance between each pair. Notice that when the sub-indexes are $i$ and $i+1$, the sum runs over all nearest successive pairs; when they are $i$ and $j$, the summation runs over all pairs including nearest successive one.

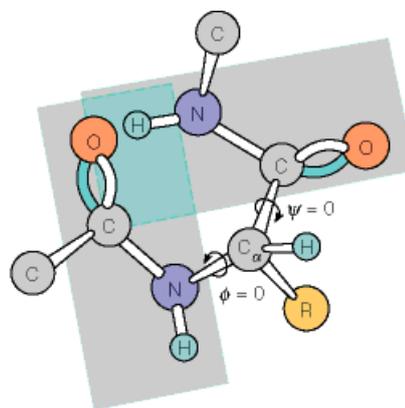

Fig. 3.7 The conformation $\phi = 0°$ and $\psi = 0°$ is not allowed in any peptides because of the steric clash between the carbonyl oxygen and amino proton. Adapted from [28].

Table 3.4 Typical minimum van der Waals radii of each atom. Adapted from [40].

| Atoms | C | H | O | N |
|---|---|---|---|---|
| Min. radius, Å | 1.5 | 1.0 | 1.3 | 1.3 |

We now specify the directional interactions. As mentioned in the beginning, water molecules are not included explicitly in our model, but their effects are incorporated into several effective potentials. $V_{directional}$ is the first one effective potential that represent the effects of water molecules. It consists of three terms,

$$V_{directional} = V_{HB} + V_{DD} + V_{DN}, \quad (3.10)$$

where $V_{HB}$ is the hydrogen bond interaction (*HB*), $V_{DD}$ is the dipole-dipole interaction (*DD*), and $V_{DN}$ is the neighboring dipole-dipole interaction (*DN*).

The hydrogen bond interaction $V_{HB}$ is a global interaction. This interaction is the most important one in our model and is the major interaction responsible for stabilizing the secondary structures [20-23]. Without this interaction, none of secondary structures can come out. In RPFM, any non-neighboring CO and NH pairs can form the hydrogen bonds by this interaction except proline. $V_{HB}$ is designed from a standard $3.6_{13}$ alpha helix with exact Ramachandran angles $\phi = -57°$ and $\psi = -47°$. It should be stressed that although the following parameters come from a standard helix structure, they could be used in the sheet conformations as well. Fig. 3.8 shows a particular relative orientation of CO and N′H′ pair when forming a hydrogen bond in helix. Obviously, $V_{HB}$ is composed of distance part and angle part. The distance part is the standard 12-10 Lennard-Jones potential form using $\sigma_{HB}$ for the equilibrium distance with $\sigma_{HB}$ being defined as the average distance between O…H′ in helix structure. As to the three angles $\theta_{1,ij}$, $\theta_{2,ij}$, and $\theta_{3,ij}$ are defined as ∠BOH′, angle between CO and N′H′, and ∠AH′O, respectively (see Fig. 3.8). These angles are employed for the relative orientation of CO and NH pairs that can form hydrogen bonds in RPFM. All of these parameters have average values as follows $\sigma_{HB} \cong 1.7834$ Å, $\theta_{1,ave} \cong 26.77°$, $\theta_{2,ave} \cong 11.60°$, and $\theta_{3,ave} \cong 17.98°$ when hydrogen bonds are forming in a standard alpha helix. Moreover, since hydrogen bond interaction is a very short range and very isotropic one, an interacting range $\theta_r$ is introduced for these three angles. Empirically, $\theta_r = 60°$. In summary, the hydrogen bond interaction in our model can be written as

$$V_{HB} = \sum_{ij} u(r_{ij}) \times v_1(\theta_{1,ij}, \theta_{1,ave}) \times v_2(\theta_{2,ij}, \theta_{2,ave}) \times v_3(\theta_{3,ij}, \theta_{3,ave}), \qquad (3.11)$$

with

$$u(r_{ij}) = \varepsilon_{HB} \times \left[ 5\left(\frac{\sigma_{HB}}{r_{ij}}\right)^{12} - 6\left(\frac{\sigma_{HB}}{r_{ij}}\right)^{10} \right] \qquad (3.12)$$

and

$$v_l(\theta_{l,ij}, \theta_{l,ave}) = \begin{cases} \left[4\cos^2(\theta_{l,ij} - \theta_{l,ave}) - 1\right] \times \frac{1}{3} & \text{when } \theta_{l,ij} < (\theta_{l,ave} + \theta_r),\ l=1,2,3 \\ 0 & \text{otherwise} \end{cases}, \qquad (3.13)$$

where $i$, $j$ represent amino acids in which $O_i$ and $H_j$ atoms are belonging to, respectively. $\varepsilon_{HB}$ is the hydrogen bond strength. $r_{ij}$ is the distance between $O_i \ldots H_j$. $\theta_{l,ij}$ are those angles defined above and $\theta_{l,ave}$ are their average values. The summation runs over all non-successive CO and NH pairs except proline.

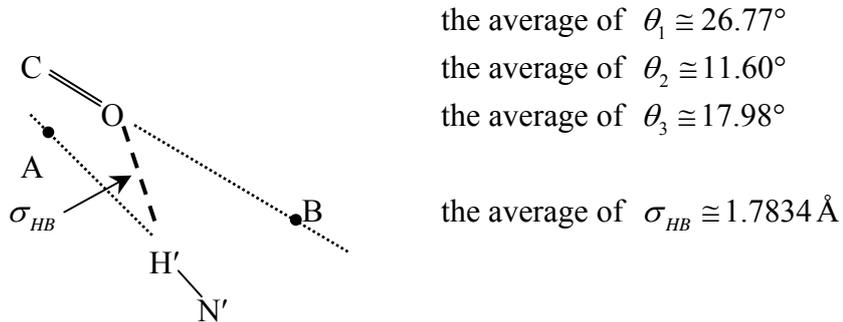

Fig. 3.8 One relative orientation of CO and NH pair when forming a hydrogen bond in a standard $3.6_{13}$ alpha helix. Point A and B are auxiliary for descriptions. $\sigma_{HB}$ is defined as the average distance between O…H′. Three angles $\theta_1$, $\theta_2$, and $\theta_3$ are defined as $\angle BOH'$, angle between CO and N′H′, and $\angle AH'O$, respectively. Their average values are in the right of this figure.

the average of $\theta_1 \cong 26.77°$
the average of $\theta_2 \cong 11.60°$
the average of $\theta_3 \cong 17.98°$

the average of $\sigma_{HB} \cong 1.7834\,\text{Å}$

The second term in the directional interactions is the global dipole-dipole interaction $V_{DD}$. This potential is one of features that make our model different from others. The idea of dipole comes from two parts. First, the CO-NH group on an amide plane has partial charges as indicated in many textbooks [38,40]. N and O atoms have an excess negative charge and H and C atoms have an excess positive charge. Because of this charge distribution, there is a dipole moment on the amide plane, as shown in Fig. 3.9. In CO-NH group, this dipole moment is almost parallel to the CO and NH bonds and is of the value $p = 1.15 \times 10^{-29}$ Cm. Moreover, when peptide chains are organized into a regular structure such as alpha helices or beta sheets, the total sum of all small dipole moment may results in a net large moment. Therefore, dipoles must have something to do with stabilizing the secondary structures. From microscopic point of view, the origin of many interactions is the dipole interaction. This is not surprising because any charge imbalance could result in dipole interactions; even the $\frac{-1}{r^6}$ part of the standard van der Waals force is a result of fluctuating dipole moments. In addition to effects of the dipole moments directly associated with the peptide, there are also indirect effects due to dipoles of water molecules, $p = 6.11 \times 10^{-30}$ Cm. For instance, dielectric constant of bulk water is about 80 which is huge in comparison to the vacuum value $\varepsilon = 1$. In general, one expects the effect of water molecules is to renormalize the strength of various interactions. For direct interactions, one would except that the dipole-dipole interaction would be one of the dominant strength. Indeed, as we shall see, while the hydrogen bonding essentially determines the helix structures. The dipole-dipole interaction is important for forming the beta sheet structures.

As argued above, we shall take $V_{DD}$ to be the standard form of dipole-dipole interaction. The overall strength is renormalized by water molecules. In RPFM, we assign two dipoles on an amide plane except proline. (Note: proline has no any dipole on its amide plane in our model). One is parallel to OC bond and at the location of O atom. The other is parallel to NH bond and at the location shifting from H atom in a small distance on the amide plane. This shifting comes from the PDB data analysis. In principle, dipole interacts between any amide planes has 4 terms (OC-OC, NH-NH, OC-NH, NH-OC). It turns out only the nearest pair dominates. Therefore, $V_{DD}$ has the form

$$V_{DD} = \varepsilon_{DD} \times \sum_{ij} \left( \frac{\mathbf{p}_i \cdot \mathbf{p}_j}{r_{ij}^3} - \frac{3(\mathbf{p}_i \cdot \mathbf{r}_{ij})(\mathbf{p}_j \cdot \mathbf{r}_{ij})}{r_{ij}^5} \right). \quad (3.14)$$

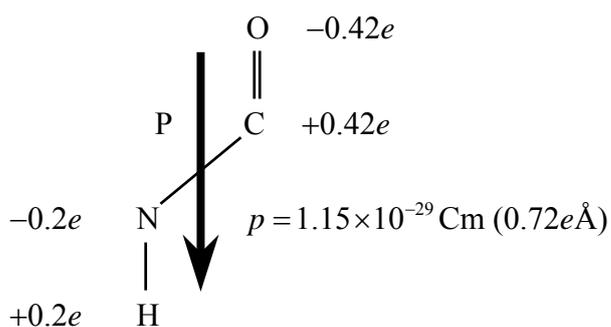

Fig. 3.9 The charge distribution and the dipole moment of an amide plane.

Here $\varepsilon_{DD}$ is the global dipole-dipole interaction strength. $\mathbf{p}_i$ and $\mathbf{p}_j$ are nearest pair of OC or NH dipoles. $r_{ij}$ is the distance between two dipoles. The sum runs over all non-successive OC and NH dipoles. Furthermore, when $\mathbf{p}_i$ or $\mathbf{p}_j$ is the OC dipole, its location is just the coordinate of O atom, while if $\mathbf{p}_i$ or $\mathbf{p}_j$ is the NH dipole, its location can be obtained by the relative coordinates of C, N, and H atoms as follows:

$$\text{The location of NH dipole} = \frac{l_2}{l_1}\left(\tilde{C} - \tilde{N}\right) + \tilde{H}, \quad (3.15)$$

where $l_1 = 0.4449$ Å and $l_2 = 1.3250$ Å. And $\tilde{C}$, $\tilde{N}$, and $\tilde{H}$ represent the coordinates of C, N, and H atom respectively.

The last directional interaction is the local dipole-dipole interaction $V_{DN}$, which is an interaction between two successive amide planes. In accordance with the orientation of successive amide planes, considering the net dipole of each peptide unit. Hence, we assign the local dipole P parallel to OC and NH bonds and at the center of backbone unit, as shown in Fig. 3.9. Since, the distance between two successive dipoles is almost constant, $V_{DN}$ has no distance dependence and would not take the form of Eq. (3.14). Instead, similar to spin-spin interaction, we assign $V_{DN}$ as

$$V_{DN} = \varepsilon_{DN} \times \sum_{i,i+1} \left( \frac{\mathbf{P}_i \cdot \mathbf{P}_{i+1}}{|\mathbf{P}_i| \times |\mathbf{P}_{i+1}|} - 1 \right) \times \frac{1}{2}, \quad (3.16)$$

where $\varepsilon_{DN}$ is the local dipole-dipole interaction strength. $P_i$ and $P_{i+1}$ are the local dipoles. The summation runs over all nearest neighboring amide planes. Notice again, proline has no this local dipole on its amide plane in our model.

Finally, we describe the hydrophobic interaction $V_{hydrophobic}$. This is the second effective potential that represents effects of water molecules in our model. Hydrophobic effect is a very important interaction induced by collective motion of water molecules. It is consider as the interaction responsible for the compact globule formation of peptides. In other words, this interaction is the major contribution to the stability of the tertiary structures. It is widely believed that the peptide will not fold definitely when this interaction disappearing. The hydrophobic interaction can be separated into global part and a local part as follows:

$$V_{hydrophobic} = V_{MJ} + V_{LocalHP} \qquad (3.17)$$

Here $V_{MJ}$ represents the global hydrophobic interaction that exists between any two non-successive side-chain units and $V_{LocalHP}$ takes care the local part of the hydrophobic interaction.

What is $V_{MJ}$? First, this is the interaction between side-chain units and therefore, a reasonable start is the so-called Miyazawa-Jernigan (MJ) matrix [26]. The MJ matrix presents the statistical behavior of side-chain units when any two of unit come close. In the literature, the MJ matrix is generalized to cover large distance interaction between any two side-chain units. The obvious extension is to use the Lennard-Jones (LJ) type potential. However, as our experience shows, such potentials would often collapse the protein immaturely. It turns out that the LJ type extension does not include effect of water molecules. To include effects of water molecules, the so-called desolvation model [31-34] was introduced.

Fig. 3.10 shows $V_{MJ}$ (black solid line) used in our model. It is obtained by combining two Gaussian functions (green dash line) into a Lennard-Jones potential (red dash line) with appropriate locations and suitable amplitudes. Here $r_1$ is the equilibrium distance between two residues in the original Lennard-Jones potential with the energy $-\varepsilon_{ij}$; $\varepsilon_{ij}$ comes from the so-called MJ matrix elements that depends on which residues are interacting. $r_2$ is the contact distance due to single water molecule. A negative Gaussian function is located at this location such that there is a small meta-stable minimum in the potential at $r_2$. In other words, two residues could be stable when there is a water molecule between them. In the middle of $r_1$ and $r_2$, i.e. $r_b$, a positive Gaussian function is located to represent the small energy cost needed for pushing away the last one water molecule between residues. This small energy cost will be termed as the desolvation barrier. All parameters in two Gaussian functions are designed such that $\frac{\varepsilon'}{\varepsilon_{ij}} \cong \frac{1}{3}$ and $\frac{\varepsilon'' - \varepsilon'}{\varepsilon' - \varepsilon_{ij}} \cong \frac{4}{3}$ in accordance with the desolvation model (see Fig. 3.10 for the meaning of $\varepsilon'$ and $\varepsilon''$). The explicit form of desolvation potential in our model is

$$V_{MJ}(r_{ij}) = \varepsilon_{MJ} \times \sum_{ij} \left[ V_{LJ}(r_{ij}) + V_{Gaussian1}(r_{ij}) + V_{Gaussian2}(r_{ij}) \right]. \tag{3.18}$$

Here $\varepsilon_{MJ}$ is the relative strength of global hydrophobic interaction. $r_{ij}$ is the distance between two $C^{\beta}$-atoms. The summation runs all over the non-successive side-chain units. $V_{LJ}$ is the original Lennard-Jones potential. $V_{Gaussian1}$ represents the positive Gaussian function, namely the desolvation barrier at $r_b$. $V_{Gaussian2}$ is the negative one which forms a small meta-stable minimum at $r_2$. The forms of these potential are

$$V_{LJ}(r_{ij}) = \varepsilon_{ij} \times \left[ \left( \frac{r_1}{r_{ij}} \right)^{12} - 2 \left( \frac{r_1}{r_{ij}} \right)^{6} \right], \tag{3.19}$$

$$V_{Gaussian1}(r_{ij}) = \varepsilon_b \times e^{-\sigma_w \times (r_{ij} - r_b)^2}, \tag{3.20}$$

$$V_{Gaussian2}(r_{ij}) = \varepsilon_2 \times e^{-\sigma_w \times (r_{ij} - r_2)^2}, \tag{3.21}$$

with

$$\begin{aligned} r_1 &= \sigma_i + \sigma_j & \varepsilon_b &= |\varepsilon''| - \varepsilon_{ij} \times \left[ \left( \frac{r_1}{r_b} \right)^{12} - 2 \left( \frac{r_1}{r_b} \right)^{6} \right] \\ r_2 &= r_1 + \sigma_w &, & \\ r_b &= \frac{r_1 + r_2}{2} = r_1 + \frac{\sigma_w}{2} & \varepsilon_2 &= -|\varepsilon'| - \varepsilon_{ij} \times \left[ \left( \frac{r_1}{r_2} \right)^{12} - 2 \left( \frac{r_1}{r_2} \right)^{6} \right] \end{aligned}, \tag{3.22}$$

and

$$\frac{\varepsilon'}{\varepsilon_{ij}} \cong \frac{1}{3}, \quad \frac{\varepsilon'' - \varepsilon'}{\varepsilon' - \varepsilon_{ij}} \cong \frac{4}{3}, \tag{3.23}$$

where $\varepsilon_{ij}$ is the MJ matrix element depending on which residues are interacting. $r_1$ is the residue-residue contact minimum with energy $-\varepsilon_{ij}$, $r_b$ is the position of desolvation barrier with energy $\varepsilon''$, and $r_2$ is the single water molecule-separated contact minimum with energy $-\varepsilon'$. $\sigma_i$ and $\sigma_j$ are the effective global radii of side-chain units in Table 3.2. $\sigma_w$ is the size of a water molecule and has the value 3Å obtained from textbooks. $\varepsilon_b$ and $\varepsilon_2$ are amplitudes of $V_{Gaussian1}$ and $V_{Gaussian2}$, respectively. These two parameters are used for rescaling the Gaussian functions such that their function values are in agreement with the desolvation model.

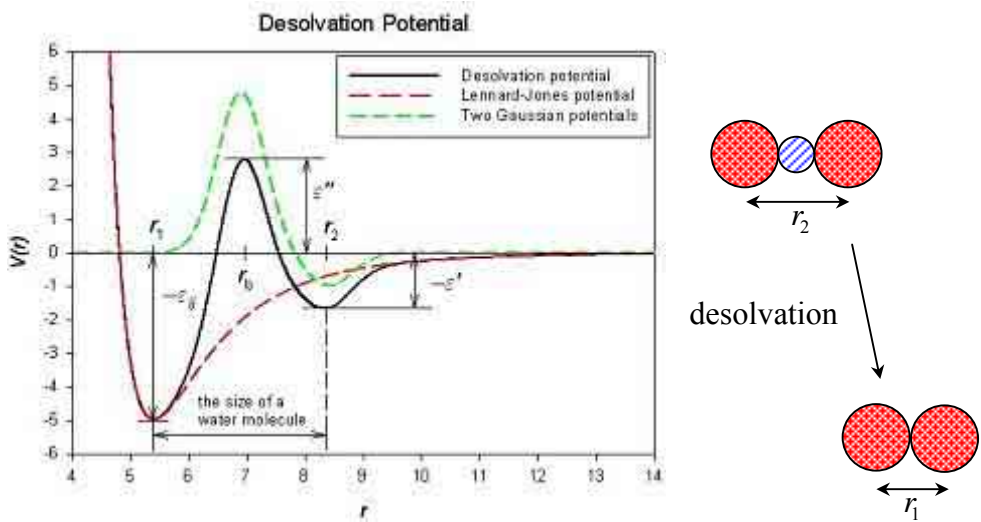

Fig. 3.10 Black solid line is the desolvation potential in RPFM. It is designed by positioning two Gaussian functions (green dash line) into a Lennard-Jones potential (red dash line) with appropriate locations and suitable amplitudes. $r_1$, $r_b$, and $r_2$ are the residue-residue contact minimum, the position of desolvation barrier, and the single water molecule-separated contact minimum, respectively. $\varepsilon_{ij}$ is the MJ matrix element depending on which residues are interacting. $\varepsilon''$ is the energy cost of the desolvation barrier. $\varepsilon'$ is the energy when there is a water molecule between two residues. These three energy parameters have the relations as follows $\frac{\varepsilon'}{\varepsilon_{ij}} \cong \frac{1}{3}$ and $\frac{\varepsilon'' - \varepsilon'}{\varepsilon' - \varepsilon_{ij}} \cong \frac{4}{3}$. In the right of this figure, the circles with big and small radius represent residues and water molecule, respectively.

The last term in the hydrophobic interaction is $V_{LocalHP}$, which is the second different feature between our model and others. This interaction is also a sequence-dependent interaction. The idea of our local hydrophobic interaction is originated from the so-called HP model and the classification of amino acids. Amino acid residues are divided into two main classes as indicated in many textbooks. Some residues that prefer to get away from water molecules are classified into the *hydrophobic* class. The others that prefer to interact with water are classified into the *polar* one. There is a tendency for hydrophobic residues to bury themselves into the interior of the protein. In contrast, the tendency is different for the polar one. They prefer exposing themselves to the water. From microscopic point of view, if two successive side-chains belonging to the same class, they tend to bury themselves inside or expose themselves on the surface of the protein. This result is as there is an attractive interaction between them. Hence, two residues in the same class have an advantage in potential when getting together. On the other hand, if two residues are not in the same class, the tendencies for them are different. It looks like there is a repulsive interaction between them. In addition, residues in the polar class can also carry charges. A new classification due to the sign of the charge on them can be introduced. When two successive side-chains have the same electrical properties, they like going to opposite direction. In contrast, they tend to get together when their charges are different. Therefore, these four charged residues are subdivided into *positive* and *negative* classes. Finally, glycine and proline are special side-chains and are classified as *neutral* class. The tendency for them is not clear. In fact, they do not participate in the local hydrophobic interaction. In summary, the local hydrophobic interaction has the form

$$V_{LocalHP} = \sum_{i,i+1} V_{XY}(r_{i,i+1}), \qquad (3.24)$$

where the sum runs all over two successive side-chain units and $r_{i,i+1}$ is the distance between them. $V_{XY}(r_{i,i+1})$ could be one of 0, $V_{HH}$, $V_{HP}$, $V_{PP}$, $V_{AT}$, or $V_{RE}$ depending on their side-chain properties and are arranged in the Table 3.5.

In Table 3.5, $V_{HH}$, $V_{PP}$, and $V_{AT}$ are interactions between two hydrophobic,

Table 3.5 Local hydrophobic interactions between any two classes. Notice that when the charged classes interact with non-charged one, they are turned into the polar class.

|  | Neutral | Hydrophobic | Polar | Positive | Negative |
|---|---|---|---|---|---|
| Neutral | 0 | 0 | 0 | 0 | 0 |
| Hydrophobic | 0 | $V_{HH}$ | $V_{HP}$ | $V_{HP}$ | $V_{HP}$ |
| Polar | 0 | $V_{HP}$ | $V_{PP}$ | $V_{PP}$ | $V_{PP}$ |
| Positive | 0 | $V_{HP}$ | $V_{PP}$ | $V_{RE}$ | $V_{AT}$ |
| Negative | 0 | $V_{HP}$ | $V_{PP}$ | $V_{AT}$ | $V_{RE}$ |

two polar, and two different charged residues, respectively. All of them have the same form that ensures the tendency for residues prefer getting together. For this reason, when two residues get closer, they gain the energy. And when they are far away, they get nothing. Interaction can be implemented by a smooth switching function and has a form of

$$V_{XY}(r) = \varepsilon_{XY} \times \begin{cases} -1 & \text{when } r < \sigma_{local} \\ \left[-1 - \cos\left(\pi \times \dfrac{r - \sigma_{local}}{0.5}\right)\right] \times \dfrac{1}{2} & \text{when } \sigma_{local} < r < \sigma_{local} + 0.5 \\ 0 & \text{when } \sigma_{local} < r \end{cases}, \quad (3.25)$$

where $\varepsilon_{XY}$ is the relative strength of local hydrophobic interaction. $XY$ means $HH$, $PP$, or $AT$ depending on the side-chain properties. $r$ is the same as $r_{i,i+1}$. $\sigma_{local}$ is the sum of the local radius of two successive side-chain units. The "0.5" is just a distance value and comes from our trial. It should be pointed out that the values of $\varepsilon_{HH}$, $\varepsilon_{PP}$, and $\varepsilon_{AT}$ could be different. The plot of Eq. (3.25) excluding $\varepsilon_{XY}$ is given in the Fig. 3.11.

On the other hand, $V_{HP}$ and $V_{RE}$ are repulsive types: $V_{HP}$ is the interaction between a hydrophobic residue and a polar one. $V_{RE}$ is the interaction between residues with the same electrical properties. Here we need to implement that when two residues are far away, they get the energy, and when they get closer, they gain nothing. For this purpose, $V_{HP}$ and $V_{RE}$ are calculated based on the angle of two vectors that are defined in two successive residues. Residues prefer far away implies that these two vectors prefer to be anti-parallel. As a result, the interaction takes the following form

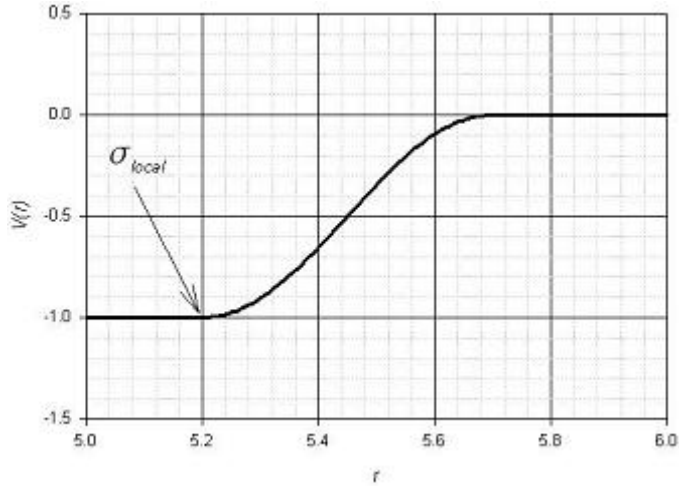

Fig. 3.11 The plot of Eq. (3.25) excluding $\varepsilon_{XY}$. The $\sigma_{local} = R_{small} + R_{small} = 5.20$ Å, for example. $V(r)$ is $-1.0$ when $r < \sigma_{local}$ and $V(r)$ is zero when $r > \sigma_{local} + 0.5$.

$$V_{XY}(r) = \varepsilon_{XY} \times \begin{cases} \left[\cos\left(\pi \times \dfrac{\theta - 90}{90}\right) - 1\right] \times \dfrac{1}{2} & \text{when } S_i \cdot S_{i+1} \leq 0 \\ 0 & \text{when } S_i \cdot S_{i+1} > 0 \end{cases}. \quad (3.26)$$

Here $\varepsilon_{XY}$ is the relative strength of local hydrophobic interaction, *XY* denotes either *HP* or *RE* depending on the side-chain properties. *r* is the same as $r_{i,i+1}$. $S_i$ and $S_{i+1}$ are vectors defined by $C^\alpha$ to the $C^\beta$-atom in *i*th and (*i+1*)th residues, respectively. $\theta$ is the angle between $S_i$ and $S_{i+1}$. Notice that the values of $\varepsilon_{HP}$ and $\varepsilon_{RE}$ could be different. The plot of Eq. (3.26) excluding $\varepsilon_{XY}$ is given in the Fig. 3.12.

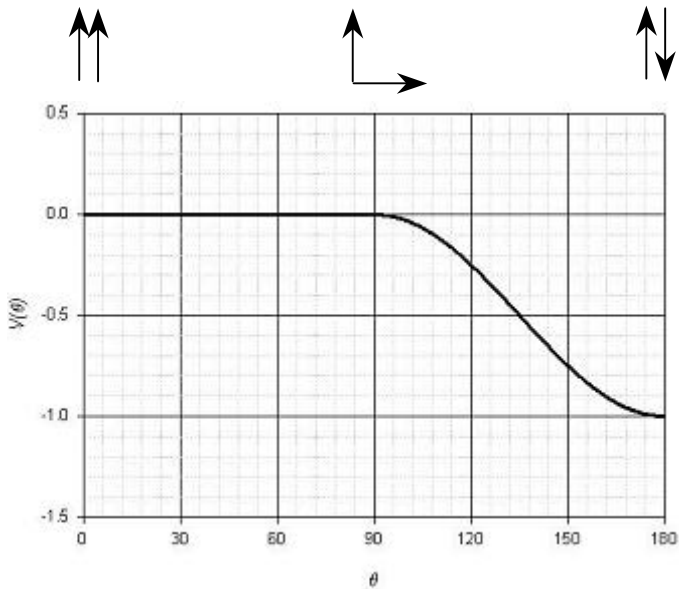

Fig. 3.12 The plot of Eq. (3.26) excluding $\varepsilon_{XY}$. The arrows above are the vectors defined by $C^\alpha$ to the $C^\beta$-atom in residues.

# §3.3 Relative Energy Strengths and Conformation Parameters

## §3.3.1 Relative Energy Strengths

In this section, we outline the method for determining relative strength of various potentials. Finally, global parameters that can be used for charactering the conformation will be introduced.

All the interactions in the previous section could be rearranged in the Table 3.6. An obvious question that remains to be addressed is how we set these relative energy strengths. These relative energy scales were denoted as $\varepsilon_{\alpha\beta}$ previously. (see Table 3.6). To determine them, we first note that, the energy reference in RPFM is $\varepsilon_{SA} = 0.6 kcal/mol$, which is the energy of room temperature. Secondly, the hydrogen bond strength is in the range of 2.4~7.2 *kcal/mol* from experiment data [28,35]. Hence, we fixed the hydrogen bond strength to the eight times of room temperature, i.e. $\varepsilon_{HB} = 4.8 kcal/mol$. The other energy strengths are less than $5.0 kcal/mol$ in principle and could be determined by the following strategy:

*Step 1*: Choose arbitrary initial value for each $\varepsilon$. Prepare a sequence in which its native conformation is already known and run folding simulations from random structures.
*Step 2*: If any conformation which energy is smaller than the native one is found, then reset the $\varepsilon$'s values and re-run the folding simulations. Repeat this step until the native conformation has the smallest energy, i.e. no other conformation encountered has smaller energy than the native one.
*Step 3*: Find another sequence and repeat *Step 2*. It should be stressed that whenever folding a new sequence, the condition in *Step 2* must be hold for all previous sequences at the same time.

Obviously, the above strategy would not yield a unique set of $\varepsilon$ parameters if only one sequence is tried. The reason is based on continuity, i.e., the energy landscape should be a continuous function of $\{\varepsilon\}$. Therefore, if a structure is located at the minimum, it will be continuously to be at the minimum for certain range of

Table 3.6 All interactions and involved units in RPFM. BB means backbone units. SC means side-chain units. $V_{SA}$, $V_{dihedral}$, $V_{HB}$, $V_{DD}$, $V_{DN}$, $V_{MJ}$, and $V_{LocalHP}$ are self-avoidance, dihedral, hydrogen bond, global dipole-dipole, neighboring dipole-dipole, global hydrophobic, and local hydrophobic interactions in the previous section, respectively. All their relative energy strengths are listed in the bottom of this table.

|  | Steric | | Directional | | Hydrophobic | |
| --- | --- | --- | --- | --- | --- | --- |
|  | Global | Local | Global | Local | Global | Local |
| BB vs BB | $V_{SA}$ | $V_{dihedral}$ | $V_{HB}, V_{DD}$ | $V_{DN}$ | × | × |
| BB vs SC | $V_{SA}$ | × | × | × | × | × |
| SC vs SC | $V_{SA}$ | × | × | × | $V_{MJ}$ | $V_{LocalHP}$ |
| Relative energy strengths | $\varepsilon_{SA}$ | $\varepsilon_{DA}, \varepsilon_{DB}$ | $\varepsilon_{HB}, \varepsilon_{DD}$ | $\varepsilon_{DN}$ | $\varepsilon_{MJ}$ | $\varepsilon_{HH}, \varepsilon_{PP}, \varepsilon_{AT}, \varepsilon_{HP}, \varepsilon_{RE}$ |

$\{\varepsilon\}$. Thus, each sequence has its own range of $\{\varepsilon\}$. If many sequences are tried, then the intersection of ranges for each sequence has to be taken. Presumably, the intersecting range will converge to a single set of $\{\varepsilon\}$. Obviously, this implies that if the above procedure is repeated again and again for various sequences, eventually a unique set of $\varepsilon$ can be found. Hence, we repeated folding simulations of some designed sequences from random structures, checked the condition in Step 2 could be hold, and made sure the native conformations could be reached for these designed sequences. Fortunately, although very time consuming, a lot of $\varepsilon$'s set could be used for these sequences. A local dominated set with the local energy larger than global one is chosen. The major reason is the screening effect of water molecules. Because of the high permittivity of water ($\approx 80$), global interactions would decay vary fast as distance increasing. Moreover, water is polar solvent. A water molecule could be treated as a small dipole as mentioned before. At the microscopic point of view, a charge could polarize the water molecules around itself. This polarization partially screens the immersed charge and diminishes the electric field in the water compared to what it would have been in a vacuum. On the other hand, local interactions interact between any two successive units. The distance between them is smaller than the size of water molecules; in other words, no any water molecule could get into the middle of two successive units. Hence, local interactions interact just like in the vacuum and the strengths of them would be larger than the global one. That is why a local dominated set is selected. Even though we select the local dominated set, the choice of $\varepsilon$ parameters set is still far from unique. This is the advantage of our model and leaves the room for folding the new sequences in the future. After a succession of folding simulation checks, we pick the $\varepsilon$ set that can fold sequence successfully as sooner as possible. This $\varepsilon$ parameters set is listed in the Table 3.7. Finally, because RPFM stores each energy component in the relative strength, monitoring the effect of each energy component is easy in our model. This will be used for analysis in the next chapter. Meanwhile, the energy landscape could be manipulated by changing these relative energy strengths to ensure the native conformation is the global minimum in energy without extensive folding simulations.

Table 3.7 The $\varepsilon$ parameters set in RPFM. The unit of all energy strengths is *kcal/mol*. The energy reference is $\varepsilon_{SA} = 0.6$, i.e. room temperature. $\varepsilon_{avoiding}$ is just a very big value so as to prevent the occurrence of unphysical situation.

| $\varepsilon_{SA}$ | $\varepsilon_{DA}$ | $\varepsilon_{DB}$ | $\varepsilon_{HB}$ | $\varepsilon_{DD}$ | $\varepsilon_{DN}$ | $\varepsilon_{MJ}$ | $\varepsilon_{HH}$ | $\varepsilon_{HP}$ | $\varepsilon_{PP}$ | $\varepsilon_{AT}$ | $\varepsilon_{RE}$ | $\varepsilon_{avoiding}$ |
|---|---|---|---|---|---|---|---|---|---|---|---|---|
| 0.6 | 1.0 | 0.5 | 4.8 | 0.2 | 2.1 | 0.2 | 5.0 | 5.0 | 5.0 | 3.0 | 3.0 | $10^6$ |

### §3.3.2 Conformation Parameters

One important problem in protein-folding problem is to find appropriate global quantities for characterizing the proteins, in particular, characterizing the conformation. In this subsection, four conformation parameters, radius of gyration ($R_G$), native contact number (*Q*), water accessible volume (*WAV*), and root-mean-square-distance (*RMSD*) will be introduced and discussed. First, the radius of gyration of a chain is defined as

$$R_G^2 = \frac{\left\langle \sum_{i=0}^{n} \rho_i^2 \right\rangle}{n+1}, \qquad (3.27)$$

where (*n*+1) is number of units and $\rho_i$ is the distance between the unit *i* and the center of gravity of the chain. Consider two units *i* and *j* at respective distances $\rho_i$ and $\rho_j$ from the center of gravity G. Because

$$r_{ij}^2 = \rho_i^2 + \rho_j^2 - 2\rho_i\rho_j \cos\phi_{ij},$$

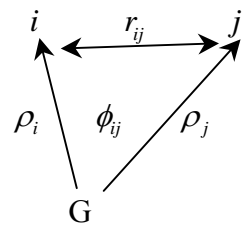

where $\phi_{ij}$ is the angle between $\rho_i$ and $\rho_j$, we obtain

$$\sum_i \sum_j r_{ij}^2 = \sum_j \left(\sum_i \rho_i^2\right) + \sum_i \left(\sum_j \rho_j^2\right) - 2\sum_i \sum_j \rho_i\rho_j \cos\phi_{ij}.$$

The last term on the right-hand side is zero by definition for the center of gravity. Since $\sum_i \rho_i^2 = \sum_j \rho_j^2 = (n+1)R_G^2$, we get

$$\sum_i \sum_j r_{ij}^2 = \sum_j (n+1)R_G^2 + \sum_i (n+1)R_G^2 = 2 \times (n+1) \times R_G^2 \times (n+1), \text{ and hence}$$

$$\Rightarrow R_G^2 = \frac{\sum_i \sum_j r_{ij}^2}{2 \times (n+1)^2} = \frac{\left(\sum_{i<j} r_{ij}^2\right)}{(n+1)^2}. \tag{3.28}$$

Thus, the radius of gyration of a chain could be obtained by calculating the distances between any two units via Eq. (3.28). This definition of radius of gyration is important since $R_G$ is a characteristic length of the molecule, which remains significant whatever the shape of the chain. In RPFM, only the backbone units are used for calculating $R_G$ and ($n+1$) is the number of backbone units.

A "contact" in RPFM is defined as when the distance between two non-successive $C^\beta$-atoms is smaller than 6.5Å, then it is called a "contact". This distance 6.5Å comes from the definition in the paper of Miyazawa and Jernigan [26]. The "native contact" is just defined as the "contact" in the native conformation. Hence, the native contact number, $Q$, can be defined by

$$Q \equiv \frac{\text{the contact the same as the native one}}{\text{the native contact}}. \tag{3.29}$$

Notice that $0.0 \leq Q \leq 1.0$. $Q = 1.0$ means peptide is in the native conformation. This definition of native contact number is also important since *Q* is an index of the structure, which indicates how similar between peptide and its native conformation.

In our study, we define a new parameter, water accessible volume (*WAV*), to further characterize energy landscape. This parameter is similar to the so-called "solvent accessible surface area" [37]. The difference is we use the volume instead of surface. Hence, *WAV* is defined as the total volume of water molecules that envelop the peptide chain within only one molecule thickness. As mentioned before, the size of a water molecule is 3Å in our model. Namely, *WAV* is just the volume of a 3Å thick envelope outside the peptide chain. Because this parameter is a continuous value, it is suitable for charactering the energy landscape with better resolution.

The root-mean-square-distance (*RMSD*) is the most common used measure, which indicates how similar between two conformations. It is defined as

$$RMSD = \sqrt{\frac{\sum_{i=1}^{n}(x_i - y_i)^2}{n}}, \quad (3.30)$$

where *n* is the number of atoms. $x_i$ and $y_i$ are the coordinates form conformation 1 and conformation 2, respectively. The definition of *RMSD* is very simple, but when calculating *RMSD*, Eq. (3.30) is useless because sometimes the orientations of two conformations may have a large difference. If calculate *RMSD* by Eq. (3.30) directly, it may cause large errors. Operationally, the *RMSD* is obtained by finding a rotation matrix M and a translated vector v such that the following quantity has minimum,

$$\sum_{i}|Mx_i + v - y_i|^2. \quad (3.31)$$

After this quantity is minimized, the *RMSD* is just the square root of Eq. (3.31). In comparing two peptide chains, only $C^\alpha$-atoms are used for calculating the root-mean-square-distance.

# Chapter 4

# Simulation Results and Analysis

In the first part of Sec. 4.1, simulation results of artificial peptides and *de novo* designed peptides are demonstrated. According to the structures, they can be classified into three different types: one alpha helix, one beta sheet, and one alpha helix and one beta sheet. Each of them has very different statistical properties and folding behavior. Their energy landscapes which show the folding properties are depicted in three different ways by using of combinations of three parameters: radius of gyration ($R_G$), native contact number ($Q$), and water accessible volume (*WAV*). The Monte Carlo evolution of the energy and these parameters are shown in this section.

    In the second part of Sec. 4.1, we analyze effects that are due to the dipole-dipole interactions and local hydrophobic interactions. These two interactions are the unique parts that distinguish our model from others. In addition, the effect of hydrogen bond interaction is also discussed in this section. We will show that hydrogen bond interaction, dipole-dipole interactions, and local hydrophobic interactions all play important roles in forming the secondary structures.

    In the final section, simulation results of real protein peptides would be given. Our model, RPFM, have folded 15 different peptides successfully. The *RMSD* for these peptides are less than 5.0Å except 4 larger peptides.

## §4.1 Simulation Results

### §4.1.1 One Alpha Helix Case

The first peptide we fold is an artificial peptide with 10 amino acids. We use four kinds of residues to design it in a periodic format. Its sequence is –Ala–Leu–Asn–Gln–Ala–Leu–Asn–Gln–Ala–Leu– (abbreviated as "ALNQ"). In RPFM, the side-chain property of this sequence has the pattern "HHPP…". Its native structure is an alpha helix with ground state energy equal to $-94.0 kcal/mol$. We fold this sequence from three different initial structures. The first initial structure is a line. All of its Ramachandran angles $\phi$ and $\psi$ are equal to $180°$. The second one is a random generating structure, which is generated by giving a random angle to each $\phi$ and $\psi$ under reasonable conditions. The third one is an unfolded structure from native conformation. It is generated by heating up the native structure slowly. All of these initial structures have zero $Q$, and no hydrogen bond is formed. All Monte Carlo simulations are running under a constant temperature $kT = 0.6$. In the following, all the simulations are done with the same setting and initial conditions.

After a series of folding simulations, all of the peptides can reach the alpha helix structures no matter what the initial structure is. The native conformations of "ALNQ" are shown in Figs. 4.1. While the corresponding native contact pairs and Ramachandran angles $\phi$ and $\psi$ are listed in Table 4.1. Because this peptide is an artificial peptide we designed, there is no standard structure to compare the *RMSD*. Simulation results show that $\phi \cong -52.68°$ and $\psi \cong -46.62°$ on the average. The $R_G$ and *WAV* of this peptide in the native state are about 5.1549Å and 5160 Å$^3$, respectively. The average number of Monte Carlo steps is around $6.2 \times 10^6$. An interesting example of run is given in Figs. 4.2, which shows the Monte Carlo evolution of the energy, radius of gyration, native contact number, and water accessible volume. All of them show dramatic collapse during the simulations. At this stage, about 70% hydrogen bond have already formed. (There are 7 hydrogen bonds in this peptide) Meanwhile, the helix structure is almost formed except two residues at each end. Because of the thermal fluctuations, it takes almost two times the steps of collapse to form the final two hydrogen bonds. This can be seen in Fig. 4.2 (a). After the collapsing, the average energy before reaching native state is about $-85.0 kcal/mol$. The difference between this value and ground state energy is exactly the energy of two hydrogen bonds. Obviously, the reason why these two hydrogen bonds form at final stage is that they are at end and thus are more loosely bound to other already-formed structure. If we ignore the bonds forming by two residues at end, it is then valid to assert that the native conformation has already been formed right after collapsing. This scenario can be seen in Figs. 4.2 (b), (c), and (d). The native contact number, $Q$, has touched 1.0, as indicated in the Fig. 4.2 (c). Meanwhile, $R_G$ and $WAV$ are almost in the constant value, as shown in the Figs. 4.2 (b) and (d), respectively. Therefore, even if the energy is far from native one, the native contact number can be used indicate whether peptide is in the native conformation or not. Finally, we note in passing that the folding curve of $R_G$ and $WAV$ have similar shapes as indicated in Figs. 4.2 (b) and (d). This implies that they are also good indicators for folding into the native structures.

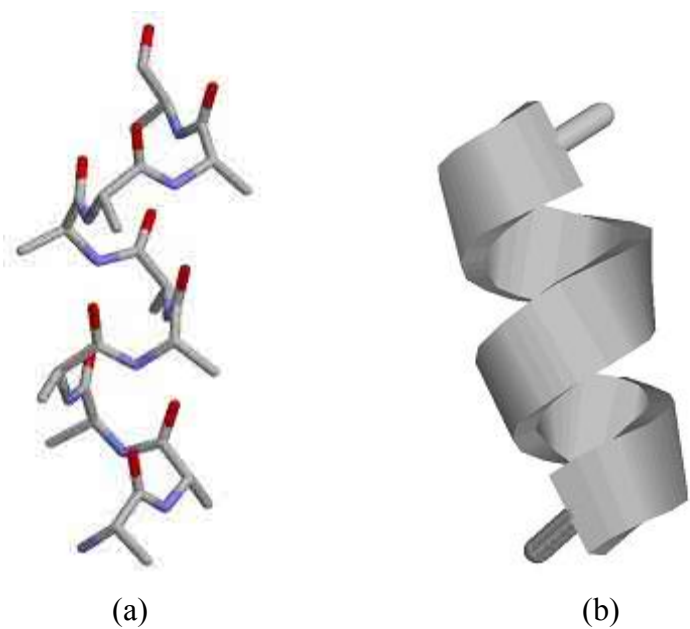

(a)                  (b)

Fig. 4.1 Native conformation of peptide "ALNQ". Left is the "Sticks" mode and right is the "Cartoons" mode in RasMol [41]. There are 7 hydrogen bonds in this structure. The ground state energy, radius of gyration, and water accessible volume are $-94.0\,kcal/mol$, 5.1549Å, and 5160 Å$^3$, respectively.

Table 4.1 Native contact pairs and Ramachandran angles $\phi$ and $\psi$ of peptide "ALNQ". The native contact map has 13 pairs. The averages of $\phi$ and $\psi$ are in the bottom of this table. Notice that the exactly alpha helix has $\phi \cong -57°$ and $\psi \cong -47°$.

| Native contact pairs | | | $\phi$, ° | $\psi$, ° |
|---|---|---|---|---|
| (1,4) | (1,5) | | −56.70 | −53.05 |
| (2,5) | (2,6) | | −53.51 | −48.94 |
| (3,6) | (3,7) | | −61.94 | −51.17 |
| (4,7) | (4,8) | | −52.89 | −47.34 |
| (5,8) | (5,9) | | −61.53 | −57.24 |
| (6,9) | (6,10) | | −49.86 | −49.92 |
| (7,10) | | | −60.31 | −50.17 |
| | | | −55.53 | −41.44 |
| | | | −68.77 | −49.75 |
| | | | −58.42 | −63.85 |
| | | Average | −52.68 | −46.62 |

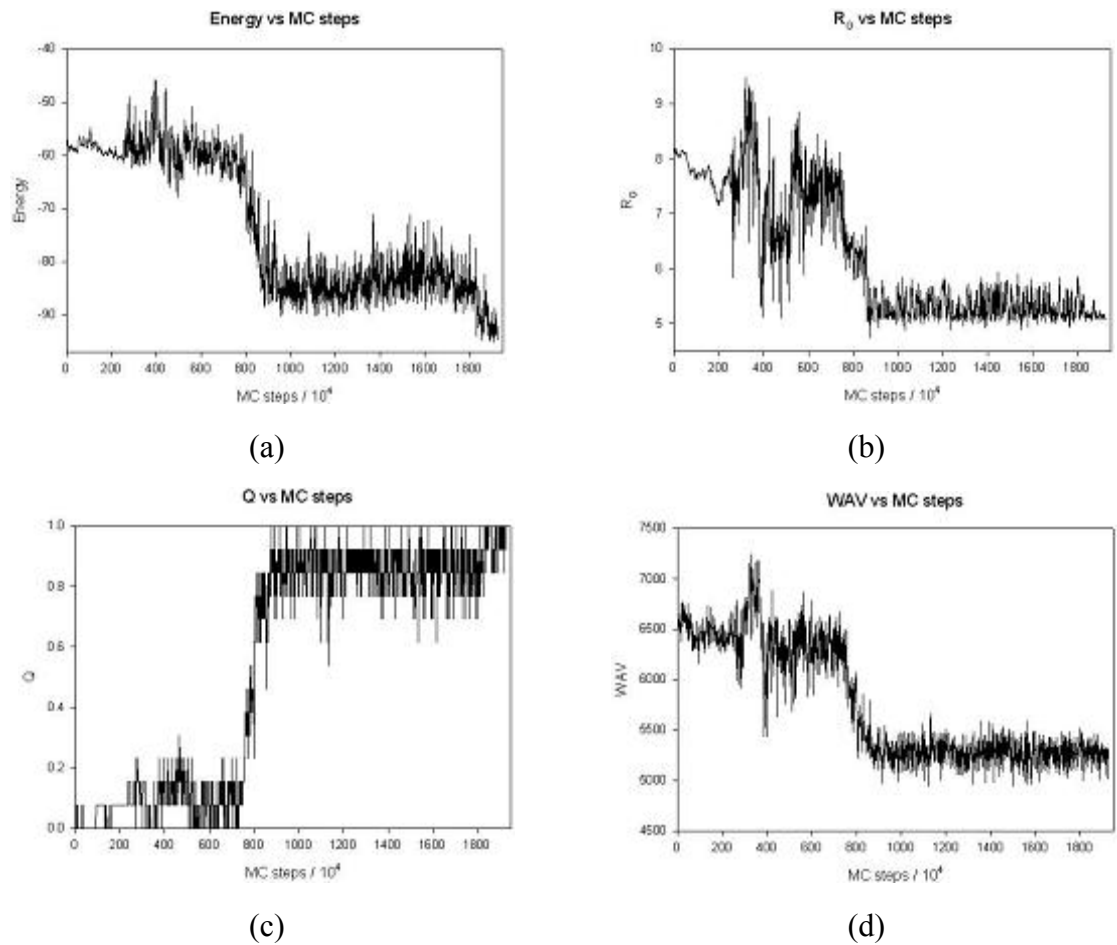

Fig. 4.2 A typical run of peptide "ALNQ" at $kT = 0.6$. Monte Carlo evolution of the energy (a), the radius of gyration $R_G$ (b), the native contact number, $Q$ (c), and the water accessible volume, $WAV$, (d). In the native state, $E = -94.0 kcal/mol$, $R_G = 5.1549$ Å, $Q = 1.0$, and $WAV = 5160$ Å$^3$.

We now turn to describe the energy landscape. The energy landscape can be depicted in three different ways by using of combinations of three parameters: $R_G$-WAV, $R_G$-Q, and WAV-Q, as shown in Fig. 4.3 (a), (b), and (c), respectively. Here energies are represented by gray scales. These figures are plotted in the way when two different conformations are in the same parameter pairs, the one with smallest energy is chosen. The initial structures are in the left half area with lower Q, larger $R_G$, and larger WAV, while the native state is the dark region at the right-bottom corner of these figures. The energy interval for the contours is $5.0 kcal/mol$. Figs. 4.3 (a), (b), and (c) indicate that the peptide "ALNQ" is almost folding smoothly from initial structure to the native state except a small barrier in Fig. 4.3 (b). In this example, this barrier is around $-85.0 kcal/mol$, which turns out to be exactly the cost of forming

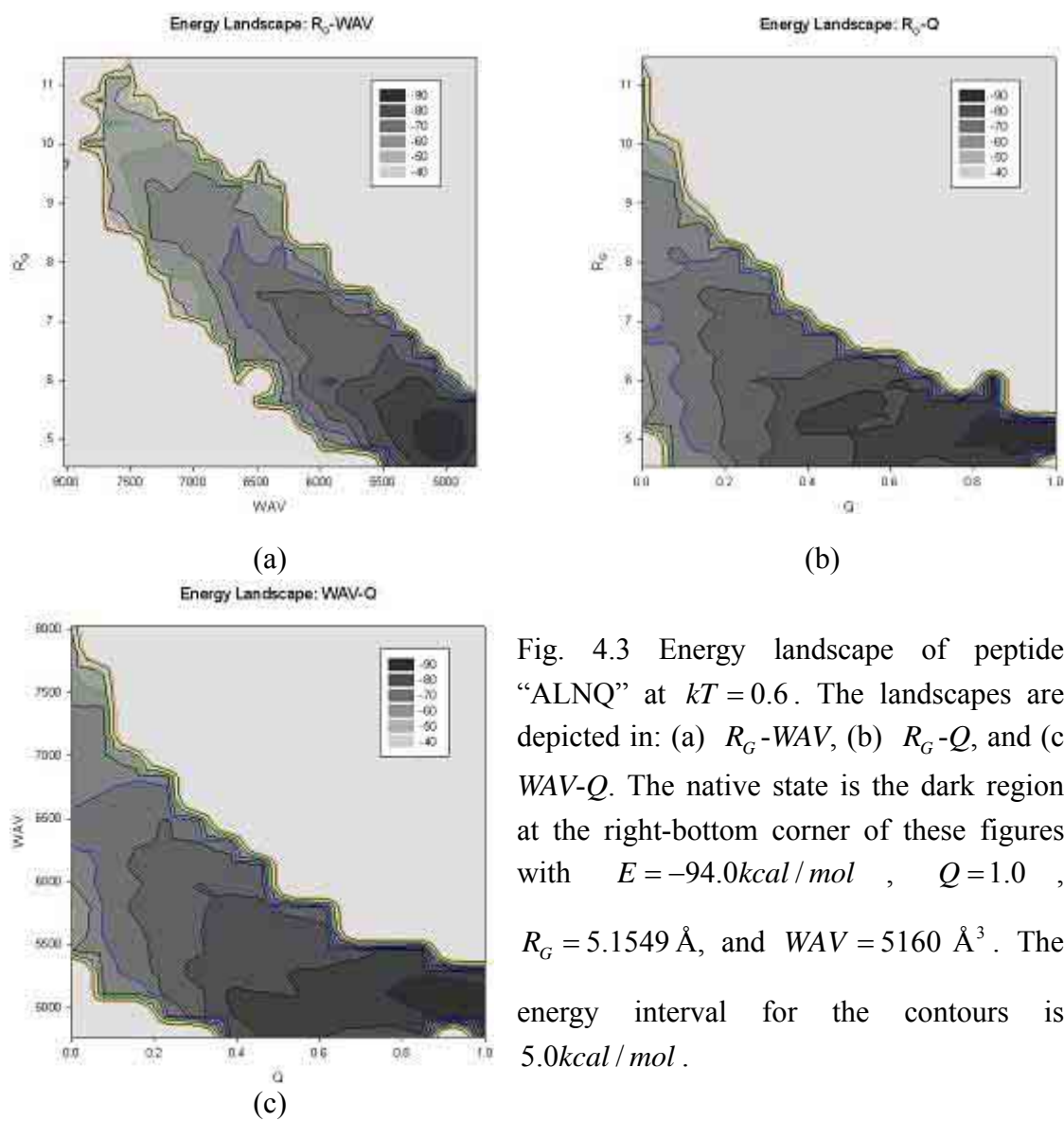

(a)　　　　　　　　　　　　　　　(b)

Fig. 4.3 Energy landscape of peptide "ALNQ" at $kT = 0.6$. The landscapes are depicted in: (a) $R_G$-WAV, (b) $R_G$-Q, and (c) WAV-Q. The native state is the dark region at the right-bottom corner of these figures with $E = -94.0 kcal/mol$, $Q = 1.0$, $R_G = 5.1549$ Å, and $WAV = 5160$ Å$^3$. The energy interval for the contours is $5.0 kcal/mol$.

(c)

final two hydrogen bonds at ends that we mentioned previously. Fig. 4.3 (a) further indicates that $R_G$ and *WAV* are good indicators for folding. The landscape contours in this figure are approximately symmetric with respect to the diagonal. It should be emphasized that because these two parameters are continuous variables, the landscape profiles in Fig. 4.3 (a) has a better resolution in comparison to (b) and (c), where *Q* is involved and it is a discrete variable. For this particular alpha, there are 13 contact pairs (see Table 4.1). Hence, *Q* has only 14 different values (0~13). When number of contact pairs decreases, more different structures would be in the same *Q*. Then, the landscape becomes more rougher and hence *Q* carries less information. Although *Q* is not a good parameter, the landscape depicted by using *Q* is more funnel-like which is broader in the initial area and narrower in the native state.

The second peptide we fold is a segment of a *de novo* designed peptide that was synthesized by DeGrado and co-workers [42]. Its motif occurs in several proteins such as myohemerythrin, apoferrittin, tabacco mosaic virus protein and cytochrome C'. Owing to its frequent occurrence in proteins, many groups have studied this peptide very detailed such as Z. Guo and D. Thirumalai [20], Shoji Takada and co-workers [21], Anders Irbäck and collaborators [22,23], and *et al*. Many excellent results have been reported in their previous studies. Here, we shall not try to analysis it in details and will only fold this segment for demonstrating that our model can do the same job. Even though only a segment of this peptide was folded, we believe that our model can fold the whole entire structures as well.

This segment has 16 amino acid residues and its sequence is –Gly–Glu–Leu–Glu–Glu–Leu–Leu–Lys–Lys–Leu–Lys–Glu–Leu–Leu–Lys–Gly– (abbreviated as "GELE"). Its side-chain property in RPFM is "$N-H--HH++H-+HH+N$". In spite of the fact that the side-chain property is different from that of "ALNQ", our model can fold it successfully. Folding simulations are done under a constant temperature $kT = 0.8$. Its native conformation is a standard alpha helix with ground state energy $-152.6 kcal/mol$. Our results show that averagely speaking, $\phi \cong -55.54°$ and $\psi \cong -58.86°$. $R_G$ and *WAV* for "GELE" in the native state are about $7.5669 Å$ and $7494 Å^3$, respectively. The average number of Monte Carlo steps is around $2.1 \times 10^7$. The Monte Carlo evolution for a typical run is shown in Figs. 4.4. Notice the stair-like behavior when energy decreases, as indicated in Fig. 4.4 (a). The energy scale of the stair is around $15.0 kcal/mol$. This corresponds to three times of the hydrogen bond energy in our model. The corresponding scenario is that after small helix turns have formed. Two small turns will come together to form a big helix structure. When this happens, three hydrogen bonds are formed at the same time. The same scenario is repeated until an entire helix conformation is formed. The above peptide has 14 hydrogen bonds and thus it is easier to see the said scenario.

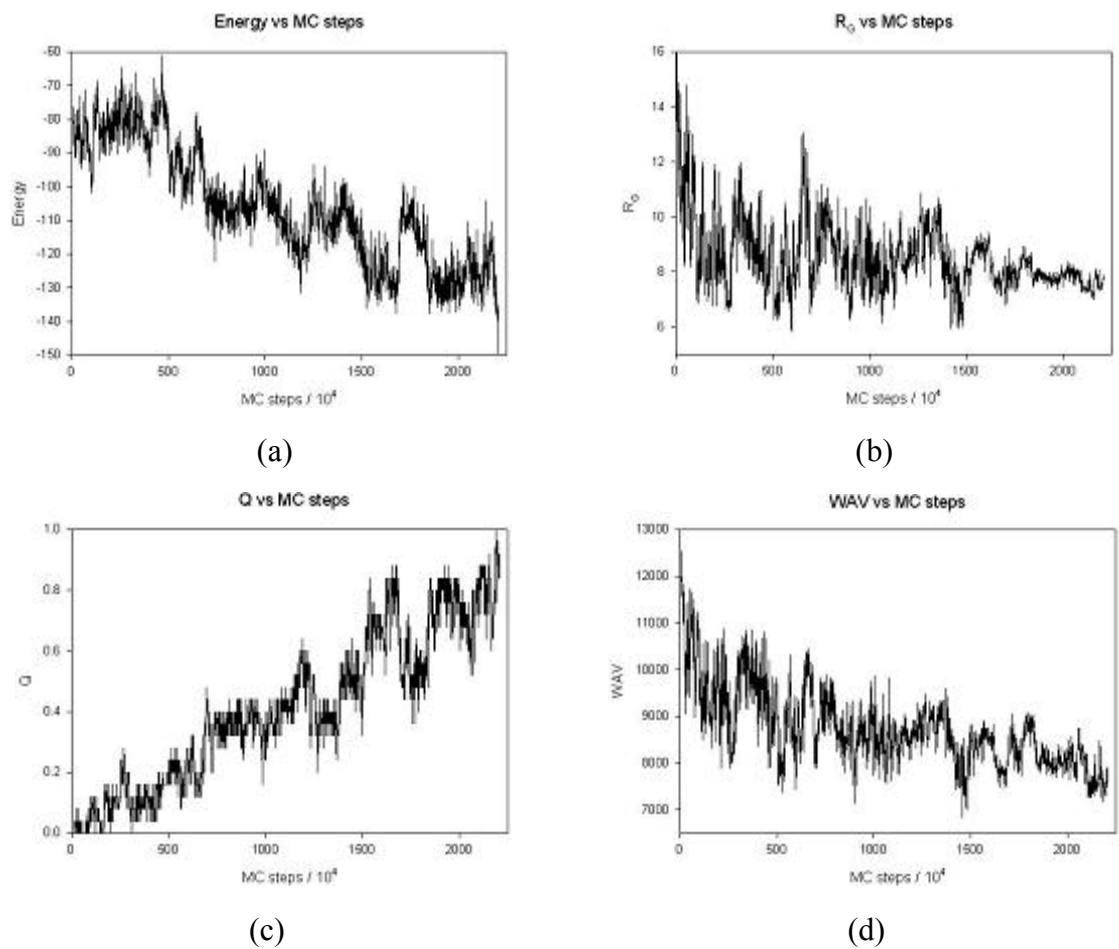

(a) (b) (c) (d)

Fig. 4.4 A typical run of peptide "GELE" at $kT = 0.8$. Monte Carlo evolution of the energy (a), the radius of gyration $R_G$ (b), the native contact number, $Q$ (c), and the water accessible volume, $WAV$, (d). In the native state, $E = -152.6 kcal/mol$, $R_G = 7.5669$ Å, and $WAV = 7494$ Å$^3$.

## §4.1.2 One Beta Sheet Case

We now turn to simulate a single beta sheet. The first structure we try to fold is also an artificial peptide with 12 residues. We use seven kinds of amino acids to design it. The sequence is –Thr–Val–Thr–Phe–Thr–Gly–Gly–Thr–Leu–Lys–Val–Tyr– (abbreviated as "TVTF"). Two glycine residues in the middle of the sequence are used for forming the beta turn. In RPFM, except the turn, the side-chain property of this sequence has the pattern of HP in alternation. We try to fold this sequence at a constant temperature $kT = 0.6$. Its native structure is a sheet with ground state energy $-97.0 kcal/mol$. The native conformations of "TVTF" are shown in Figs. 4.5 and the native contact pairs and Ramachandran angles $\phi$ and $\psi$ are listed in Table 4.2.

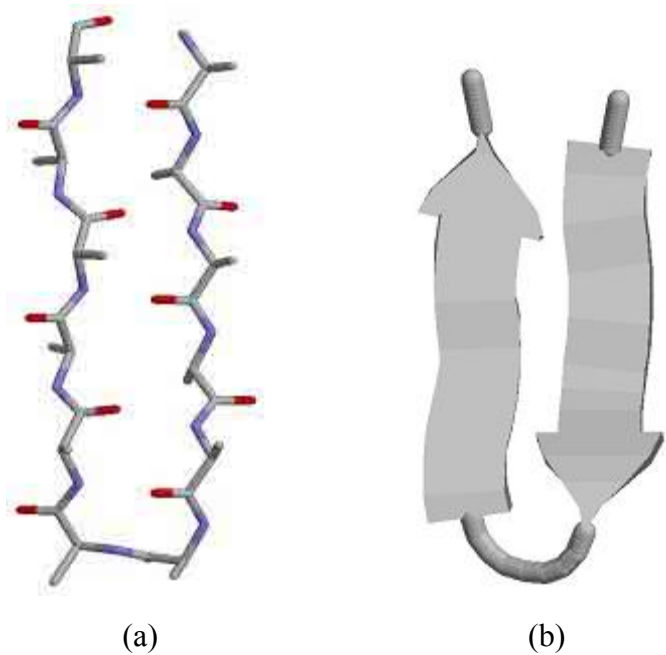

(a)                  (b)

Fig. 4.5 Native conformation of peptide "TVTF". Left is the "Sticks" mode and right is the "Cartoons" mode in RasMol. There are 6 hydrogen bonds in this structure. The ground state energy, radius of gyration, and water accessible volume are $-97.0 kcal/mol$, 6.8034Å, and 6265 Å$^3$, respectively.

Table 4.2 Native contact pairs and Ramachandran angles $\phi$ and $\psi$ of peptide "TVTF". The native contact map has 6 pairs. The averages of $\phi$ and $\psi$ except the turn are in the bottom of this table. Notice that the exactly anti-parallel beta sheet has $\phi \cong -139°$ and $\psi \cong +135°$.

| Native contact pairs | | $\phi$, ° | $\psi$, ° |
|---|---|---|---|
| (1,12) | | −130.87 | +145.76 |
| (2,11) | | −137.99 | +124.66 |
| (3,10) | | −130.99 | +128.75 |
| (4,9)  | | −129.21 | +131.86 |
| (5,8)  | | −147.92 | +140.35 |
| (6,8)  | | +37.71  | −107.29 |
|        | | −117.67 | +44.27  |
|        | | −143.61 | +129.51 |
|        | | −125.88 | +125.41 |
|        | | −127.60 | +116.81 |
|        | | −126.30 | +131.04 |
|        | | −136.39 | +133.28 |
|        | Average | −133.68 | +130.74 |

Our results show that $\phi \cong -133.68°$ and $\psi \cong +130.74°$ when two angles at turn are excluded. $R_G$ and *WAV* of this peptide in the native state are about 6.8034Å and 6265 Å$^3$, respectively. The average number of Monte Carlo steps is about $7.1 \times 10^7$ which is an order of magnitude larger than that for the helix. A typical result is given in Figs. 4.6, which shows the Monte Carlo evolution of the energy, $R_G$, *Q*, and *WAV*. All of them also show the same collapse in the middle of running as that in the helix case. The collapse corresponds to the moment when two arms of the beta sheet contact each other. At that moment, the sheet structure is almost formed except two hydrogen bonds at ends of the arms. After the collapsing, the average energy before reaching the native state is about $-87.0 kcal/mol$. The difference between this value and the ground state energy is again energy of two hydrogen bonds, as shown in Fig. 4.6 (a). If we ignore these two bonds, peptide has formed the native conformation after collapsing. It is interesting to note that Figs. 4.6 (b) and (d) show that both $R_G$ and *WAV* first collapse in the early stage and then come back to high values. When they collapse, energy only drops slightly and Q increases to 0.66 approximately, as indicated in Figs. 4.6 (a) and (c). This is because two arms have contacted to each other, but due to their misorientation, hydrogen bond cannot be formed. Therefore, two arms reopen and continue to find their native conformations. Since there is no hydrogen bond formed between two arms, the duration for reopening is short.

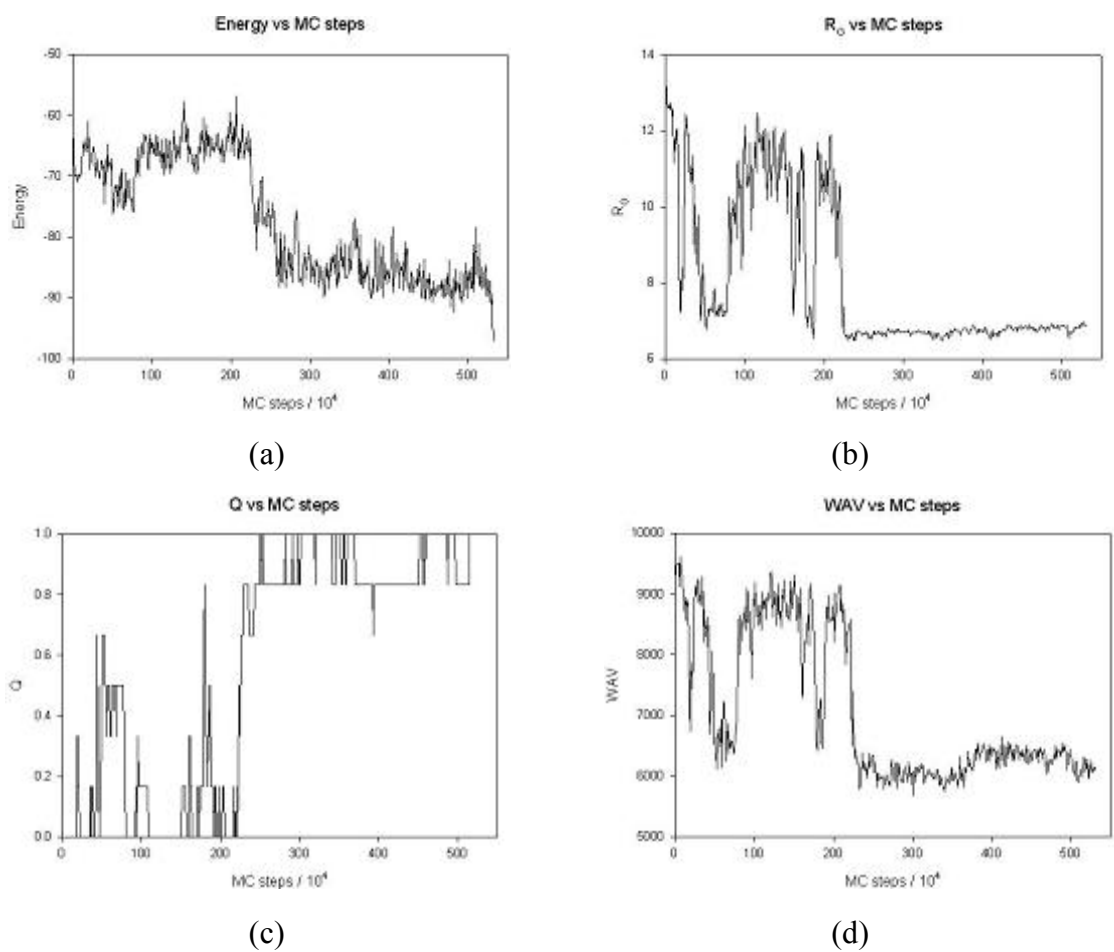

Fig. 4.6 A typical run of peptide "TVTF" at $kT = 0.6$. Monte Carlo evolution of the energy (a), the radius of gyration $R_G$ (b), the native contact number, *Q* (c), and the water accessible volume, *WAV*, (d). In the native state, $E = -97.0 kcal/mol$, $R_G = 6.8034$ Å, $Q = 1.0$, and $WAV = 6265$ Å$^3$.

In the above simulations, the hydrogen bonds of the beta sheet structure are formed by starting from the turning place and the then moving to the end to end. However, some simulation results in the literature show the opposite sequence, i.e. folding starts from the end and moves to the turning place. The probability of the latter is low because the probability for two ends of the arm to contact each other in right orientation for forming the hydrogen bonds is low. Once two arms contact in the right orientation, the hydrogen bonds will be formed in succession very quickly. Then the peptide can reach the native state faster than that starts from the turning place.

Note that two arms in this beta sheet might mismatch when hydrogen bonds form (see Fig. 4.7). For short peptides, this mismatch shifts one hydrogen bond up. In RPFM, the forming of hydrogen bonds between two arms often starts from the turning point. Two backbone units near the turn have large chance to form the hydrogen bonds. If shifting happens near the turn in the beginning, the succeeding hydrogen bonds will be all wrong. Hence, when mismatch occurs, it is necessary to break all the wrong bonding and try again. This is the reason why beta sheet folds slower than the alpha helix. On the other hand, if folding starts from the end, since two backbone units at the end have small chance to collide in the appropriate orientation for the hydrogen bond, the mismatch probability is low. We should emphasize that the situation would be different for the long beta sheet peptide. In this case, because shift of more than one hydrogen bond can occur, it turns out both cases can produce mismatch. Therefore, long beta sheet structures are more difficult to fold.

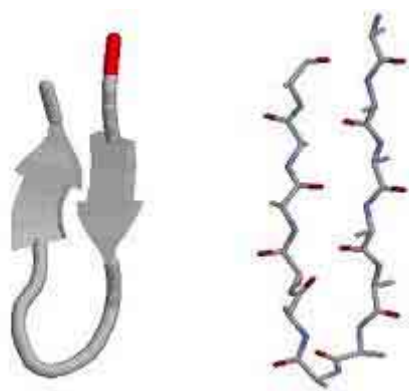

Fig. 4.7 Mismatch in forming of hydrogen bonds. The energy, $R_G$, Q and *WAV* of this conformation are about $-92.4 kcal/mol$, 6.6Å, 0.33, and 6300 Å$^3$, respectively.

We now turn to describe the energy landscape. Again, energy landscapes are depicted in three ways: $R_G$-*WAV*, $R_G$-*Q*, and *WAV*-*Q*, as shown in Figs. 4.8 (a), (b), and (c), respectively. The initial structures are in the left half area with lower *Q*, larger $R_G$, and larger *WAV*. The native state is the dark region at the right-bottom corner of these figures. The energy interval for the contours is $5.0 kcal/mol$. Energy landscape profiles of "TVTF" are more interesting than the helix case. First of all, roughly speaking, when the peptide collapses, the decreasing of $R_G$ happens earlier than the decreasing of *WAV*, as indicated in Fig. 4.8 (a). This implies that two arms of the sheet get together first, then they push away the water molecules between them. As mentioned in the previous chapter, pushing away the water molecules will cost energy. This energy cost corresponds to the desolvation barrier in desolvation model. Both in the Figs. 4.8 (a) and (b) shows these desolvation barriers clearly (indicated by red arrows in these figures). Secondly, there is a basin at *Q* around 0.4, as indicated by green arrows in Figs. 4.8 (b) and (c), and $R_G$ and *WAV* corresponding to this basin are around 6.3~6.7Å and 5800~6400 Å$^3$, respectively. This basin is exactly the location of the mismatch conformation as indicated from the conformation parameters (mismatch conformation has $R_G = 6.6$ Å, $Q = 0.33$, and $WAV = 6300$ Å$^3$).

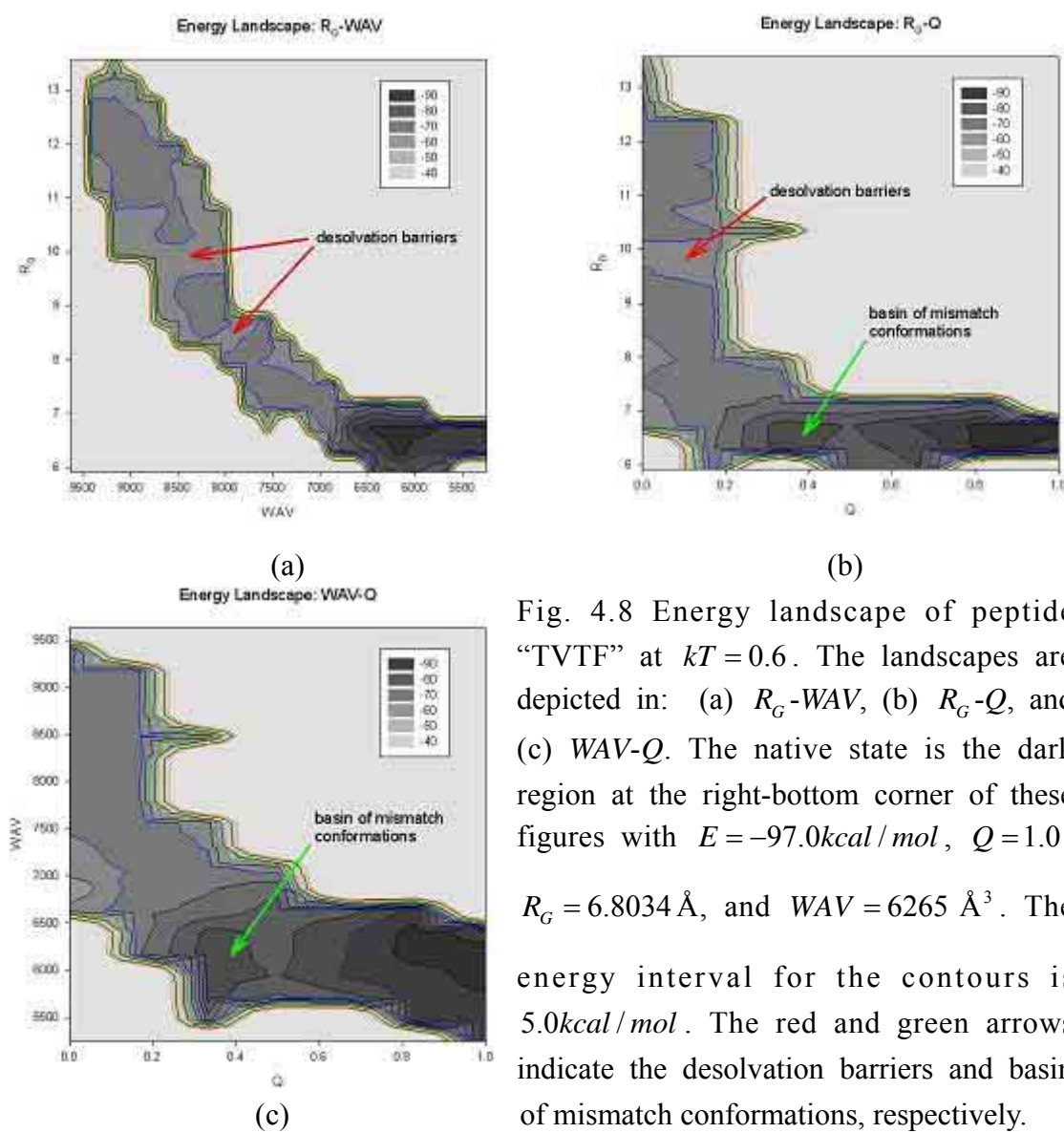

Fig. 4.8 Energy landscape of peptide "TVTF" at $kT = 0.6$. The landscapes are depicted in: (a) $R_G$-*WAV*, (b) $R_G$-*Q*, and (c) *WAV*-*Q*. The native state is the dark region at the right-bottom corner of these figures with $E = -97.0 kcal/mol$, $Q = 1.0$, $R_G = 6.8034$ Å, and $WAV = 6265$ Å$^3$. The energy interval for the contours is $5.0 kcal/mol$. The red and green arrows indicate the desolvation barriers and basin of mismatch conformations, respectively.

Meanwhile, this basin occupies almost half of the channel between initial structure to the native state, as illustrated in Figs. 4.8 (b) and (c). Hence, when peptide folds from initial structure to the native one, it has a large chance to get into this basin, i.e., peptide folds to the mismatch conformation. This explains again why the beta sheet is easily folding to the mismatch conformation and is slower than the helix. Why is the basin missing in Fig. 4.8 (a)? This is because when two structures are in the same conformation parameter pairs, only the lowest energy will be selected by our definition. The values of $R_G$ and *WAV* for the mismatch conformation are very close to those of the native one. Since only the conformation with lowest energy, i.e., the native state, is chose; hence the mismatch basin disappears in this figure. However, as for *Q*, mismatch conformation and native one have different *Q* values. Thus, this basin can be seen both in Figs. 4.8 (b) and (c). Finally, we want to address why the desolvation barrier only appears in the beginning. This is because after two arms in the beta sheet are contacted each other; there is no water molecule between them near the native state. Since after two arms contact each other, the only thing that the peptide does is to adjust their local orientations to form the hydrogen bonds more precisely. As a result, the action of pushing water only occurs at the beginning.

## §4.1.3   One Alpha Helix and One Beta Sheet Case

To further test our model, we turn to fold a peptide with both helix and sheet structures. The first peptide with both helix and sheet is also an artificial peptide with 24 residues. It is designed by the combination of two sequences in the previous two sections with some modification. Its sequence is –Thr–Ala–Thr–Leu–Gly–Gly–Val–Lys–Ala–Tyr–Gly–Gly–Asn–Gln–Ala–Leu–Asn–Gln–Ala–Leu–Asn–Gln–Ala–Leu–

(abbreviated as "TATL"). The first two glycines are used for forming the beta turn in the middle of sheet structure and the next two glycines are used for connecting sheet and helix structures. The first 10 and last 12 residues in this peptide are responsible for forming sheet and helix conformations, respectively. This peptide is folded at a constant temperature $kT = 0.8$. Its native state energy is $-204.0 kcal/mol$. The native conformations and native contact map of "TATL" are shown in Figs. 4.9 (a) and (b), respectively. Its Ramachandran angles $\phi$ and $\psi$ are listed in Table 4.3. Our results show that $R_G$ and $WAV$ of this peptide in the native state are around 7.1422Å and 8276 Å$^3$, respectively. The average number of Monte Carlo steps is $1.2 \times 10^8$ approximately. Meanwhile, the average values of $\phi$ and $\psi$ for the sheet part are $-131.70°$ and $+129.81°$, respectively. These two values exclude the angles of the beta turn. For the helix part, $\phi \cong -53.76°$ and $\psi \cong -56.33°$, respectively.

As a result of two different structures in this peptide, its folding process is more complicated than the previous one. Most of our simulation results show that the major folding scene of this peptide is helix structure formed before the beta sheet. The Monte Carlo evolution of the energy and $Q$ show this process clearly, as indicated in

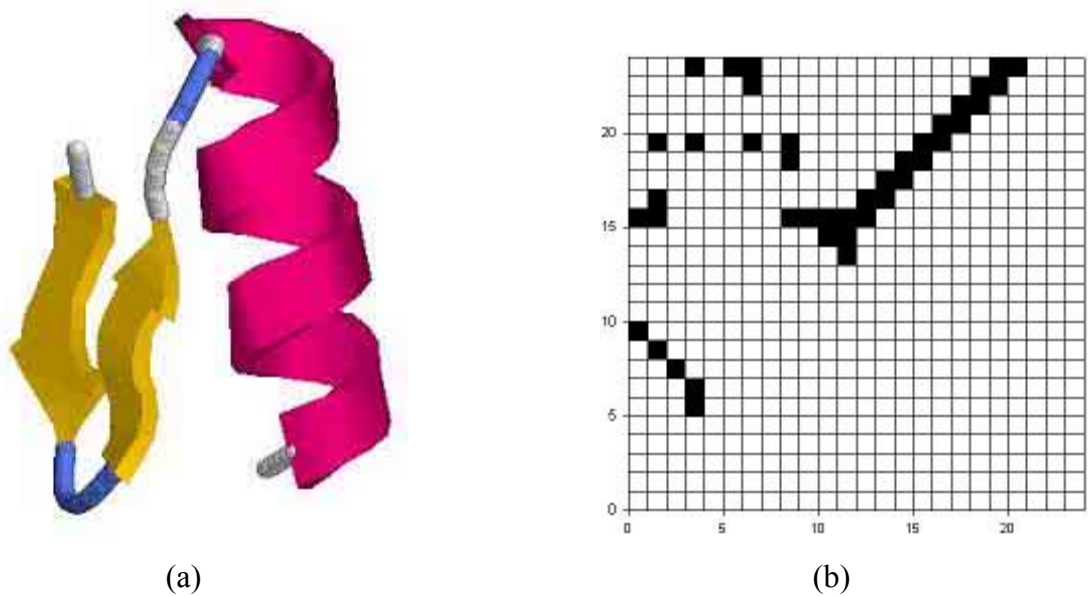

(a)                                                              (b)

Fig. 4.9 (a) Native conformation of peptide "TATL" drawn in the "Cartoons" mode in RasMol. Yellow, pink and blue parts are sheet, helix and turn structures, respectively. Its native state energy, $R_G$, and $WAV$ are $-204.0 kcal/mol$, 7.1422Å and 8276 Å$^3$, respectively. (b) Native contact map of "TATL". This peptide has 41 contact pairs as indicated by the black squares.

Table 4.3 Ramachandran angles $\phi$ and $\psi$ of peptide "TATL". The average values of $\phi$ and $\psi$ for sheet and helix structures are in the bottom of this table, respectively. The average values for the sheet part are excluding the angles of the beta turn in the middle of the sheet structure. Notice that the exactly alpha helix has $\phi \cong -57°$ and $\psi \cong -47°$. And the exactly anti-parallel beta sheet has $\phi \cong -139°$ and $\psi \cong +135°$.

| Sheet structure | | Turn between sheet and helix | | Helix structure | |
|---|---|---|---|---|---|
| $\phi$, ° | $\psi$, ° | $\phi$, ° | $\psi$, ° | $\phi$, ° | $\psi$, ° |
| −143.52 | +124.02 | +154.91 | −72.83 | −56.22 | −62.22 |
| −125.01 | +138.38 | −147.07 | +128.13 | −53.29 | −53.79 |
| −139.46 | +111.92 | −26.20 | −53.39 | −51.18 | −55.81 |
| −139.28 | +131.87 | | | −54.98 | −60.07 |
| −57.33 | −101.11 | Beta | | −51.52 | −63.61 |
| −127.65 | +15.00 | Turn | | −47.74 | −48.19 |
| −110.57 | +131.12 | | | −57.04 | −63.38 |
| −130.36 | +121.65 | | | −48.84 | −44.43 |
| −119.47 | +142.54 | | | −61.95 | −65.62 |
| −145.92 | +136.97 | | | −47.53 | −44.82 |
| −131.70 | +129.81 | Average | | −61.06 | −57.77 |
| | | | Average | −53.76 | −56.33 |

Figs. 4.10 (a) and (b). In the first $10^7$ steps, $Q$ increases from 0.0 to 0.5 and energy decreases from $-100.0 kcal/mol$ to $-160.0 kcal/mol$ in a small stair-like behavior. As mentioned in the previous section, this stair-like behavior indicates the helix structure is constructed at this stage. After this stage, helix structure is formed as indicated by the $Q$ value (In this peptide, $Q$=0.5 corresponds to the case when only the helix structure is formed.) and the snapshots above the Fig. 4.10 (a). In the next $10^7$ steps, $Q$ and energy are almost fixed. This indicates that the helix part maintains its structure, while the remaining part has no specific structures. It also can been seen in the large variations in the Monte Carlo evolution of $R_G$ and *WAV*, as indicated in Figs. 4.10 (c) and (d). In the final stage, i.e. after the $2.2 \times 10^7$ Monte Carlo steps, Fig. 4.10 (a) shows a dramatic collapsing. This corresponds to the forming of sheet structure. After the sheet is formed, the helix and sheet structures may not be in the right orientation. Hence, these two domains will try to find their right orientation by repeatedly opening and collapsing in the remaining steps. Finally, peptide completes the folding of tertiary structures and gets into the native conformation. Some snapshots of these processes are also given in the top of Fig. 4.10 (a). We note in passing that this is the clearest run that shows the folding of helix and sheet structures in the Monte Carlo evolution explicitly. In most of the other runs, these features are not that clear in the evolution.

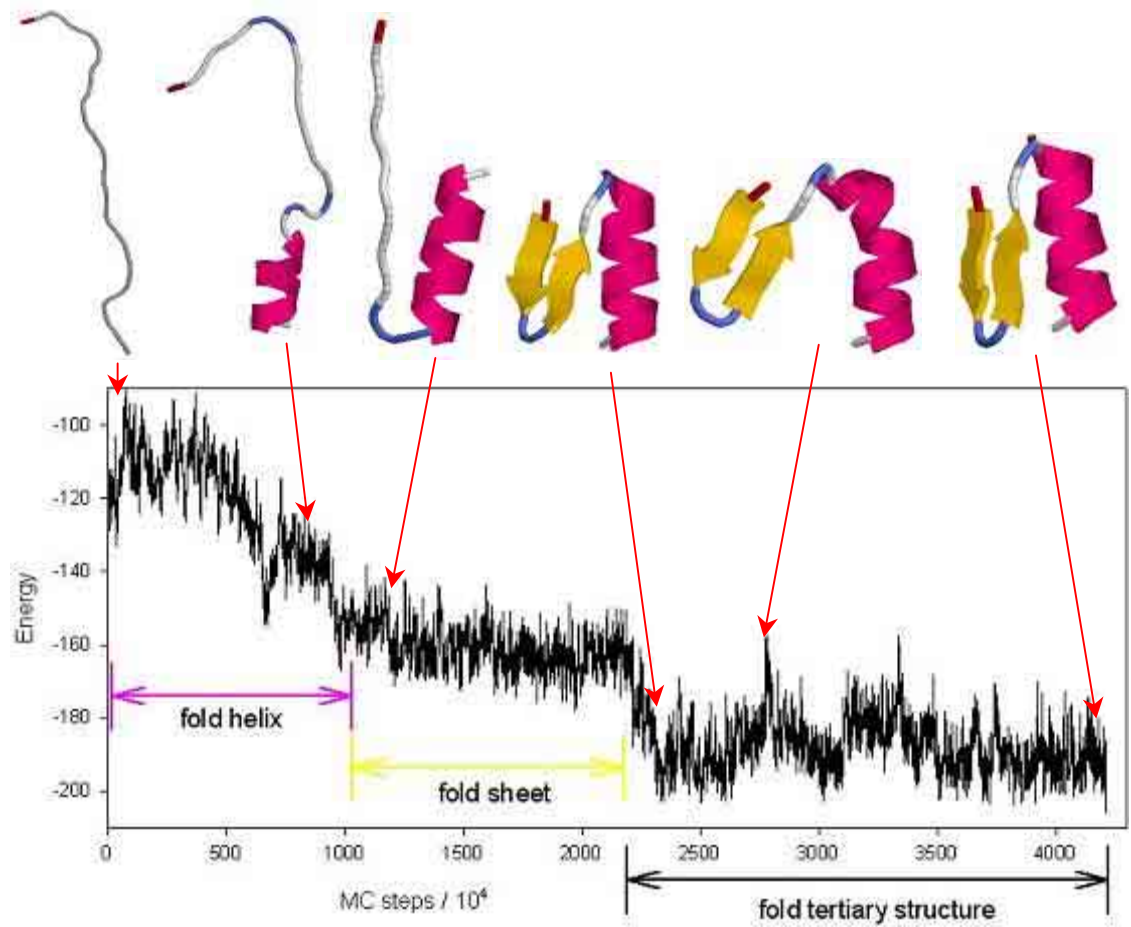

(a)

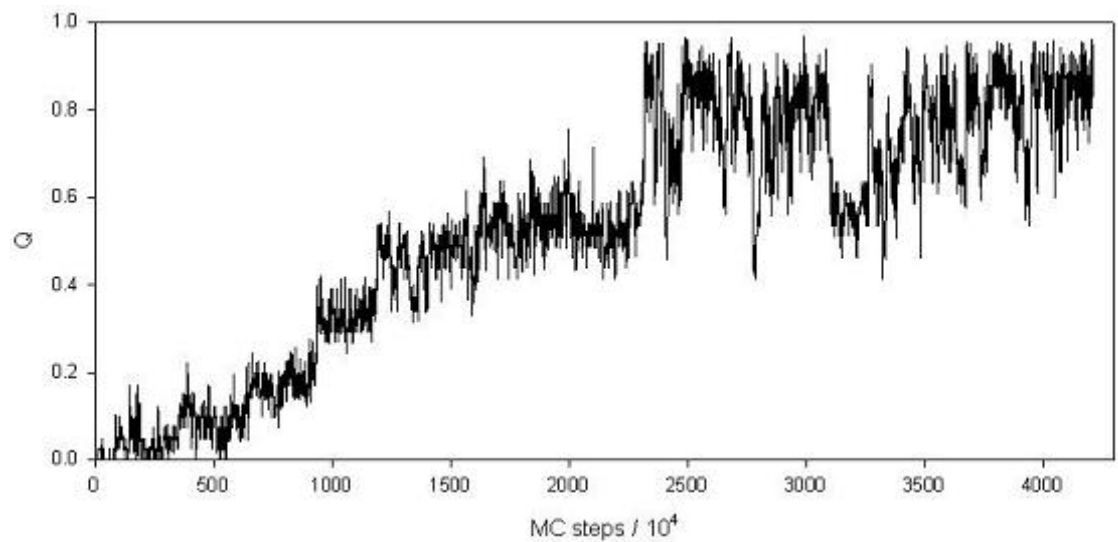

(b)

Fig. 4.10 An example of folding the peptide "TATL" at $kT = 0.8$. Monte Carlo evolution of the energy (a) and the native contact number, Q (b). In the native state, $E = -204.0 kcal/mol$. Snapshots are in the top of these figures.

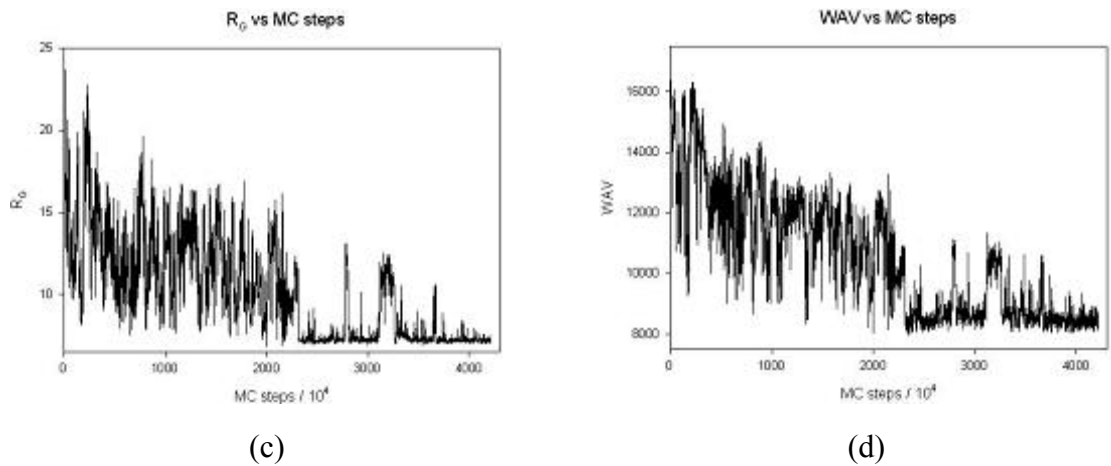

(c)                            (d)

Fig. 4.10 An example of folding the peptide "TATL" at $kT = 0.8$. Monte Carlo evolution of the radius of gyration $R_G$ (c) and the water accessible volume, $WAV$ (d). In the native state, $R_G = 7.1422$ Å, and $WAV = 8276$ Å$^3$.

    Finally, let us turn to depict the energy landscape for this peptide. They are shown in Figs. 4.11 (a), (b), and (c). The initial structures are in the left half area with lower $Q$, larger $R_G$, and larger $WAV$. The native state is the dark region at the right-bottom corner of these figures. The energy interval for the contours is $5.0 kcal/mol$. There are three color circles both in Figs. 4.11 (b) and (c). These circles are used for indicating the location of different conformations in the Fig. 4.12. The location of yellow circles correspond to the conformations which only helix is formed and sheet is not yet, as indicated in Fig. 4.12 (a). The location of red circles correspond to the conformations which helix and sheet are all formed, but they are not compact enough yet, as shown in Fig. 4.12 (b). Finally, the location of blue circles correspond to the conformations which helix and sheet are all formed, but they are packed in the wrong orientation, as illustrated in Fig. 4.12 (c). Notice that as a result of too many conformations occurring for the same parameter pairs, the energy landscape figures are highly coarse grained.

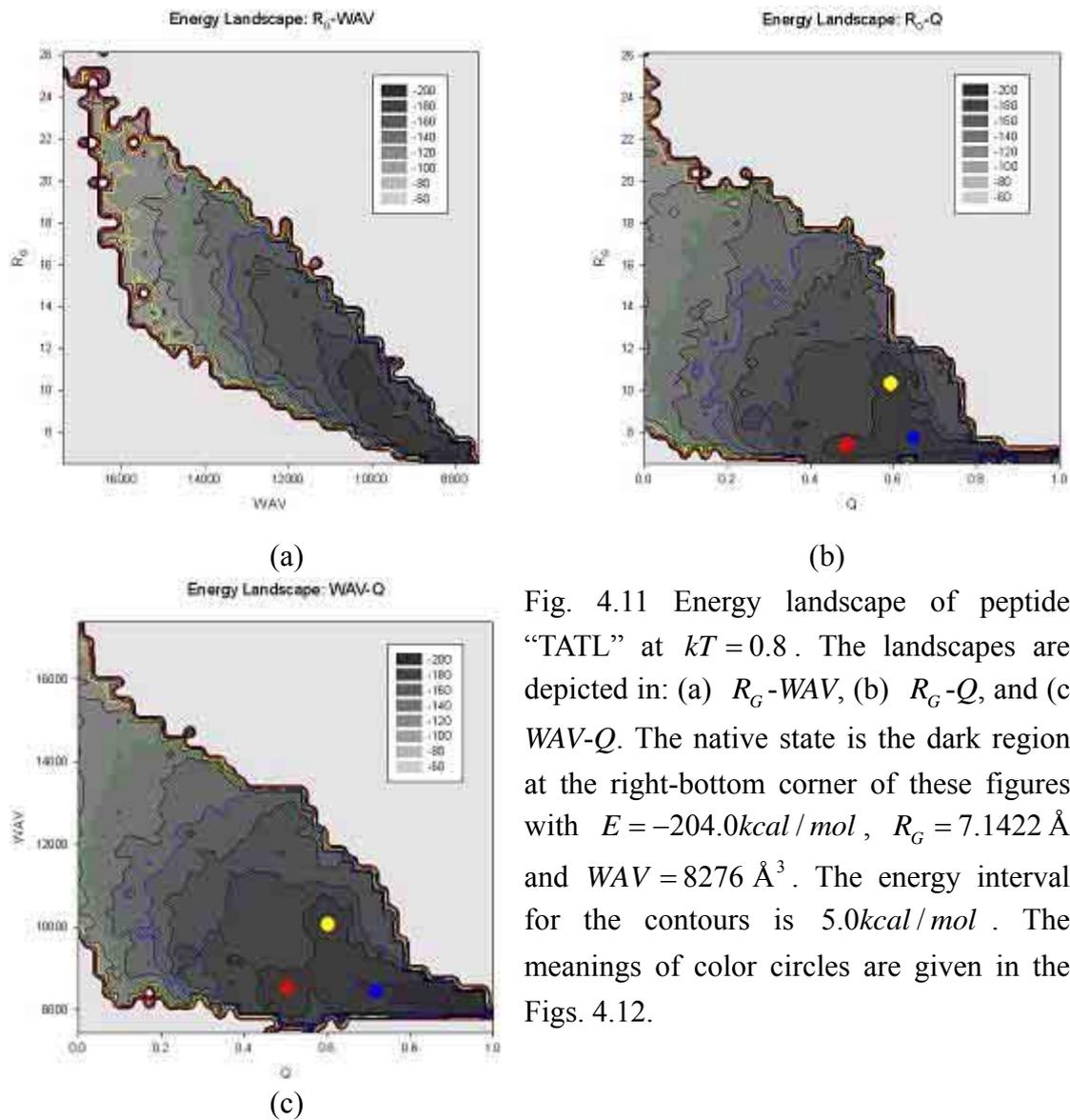

Fig. 4.11 Energy landscape of peptide "TATL" at $kT = 0.8$. The landscapes are depicted in: (a) $R_G$-*WAV*, (b) $R_G$-*Q*, and (c) *WAV*-*Q*. The native state is the dark region at the right-bottom corner of these figures with $E = -204.0 kcal/mol$, $R_G = 7.1422$ Å, and $WAV = 8276$ Å$^3$. The energy interval for the contours is $5.0 kcal/mol$. The meanings of color circles are given in the Figs. 4.12.

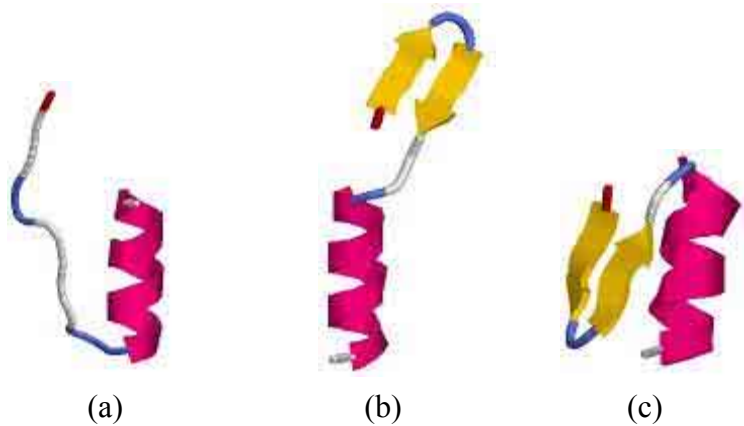

Fig. 4.12 The yellow, red, and blue circles in Figs. 4.11 (a) and (b) correspond to the location of approximate conformations of Figs. 4.12 (a), (b), and (c), respectively.

## §4.1.4 Effects of Dipole-Dipole Interactions and Local Hydrophobic Interactions

In this subsection, we analyze effects that are due to the dipole-dipole interactions and local hydrophobic interactions. These two interactions are the unique parts that distinguish our model from others. For this purpose, we systematically decrease the strength of these two interactions separately and re-run the simulations. The effects of these two interactions can be seen in the one-dimensional energy landscapes along the folding pathway. Due to the fact that conformations in the same $R_G$ or *WAV* may belong to two totally different structures, we shall only present the energy landscapes in the parameter $Q$ so that $Q=1.0$ corresponds to the unique native conformation with certain.

First, we consider the effect of dipole-dipole interactions. Figs. 4.13 (a) and (b) show the effect of dipole-dipole interactions of three different strength (indicating by parameter DD) for the one helix case and one beta sheet case, respectively. We can see that the native conformation for the helix is not affected when the strength of the dipole-dipole interactions changes. The helix peptide still reaches the native state even the dipole-dipole interactions are turned off. The landscape profiles only change slightly as the interactions strength decreases. On the contrary, the dipole-dipole interactions do affect the beta sheet structures. When the strength decreases to the half of the original one, the native state still the same, but the landscape profile deforms violently. When the dipole-dipole interactions are turned off, the original native state for the sheet conformation is no longer the ground state. Hence, the dipole-dipole interactions have strong effects on the beta sheet structure, while it does not affect the helix one.

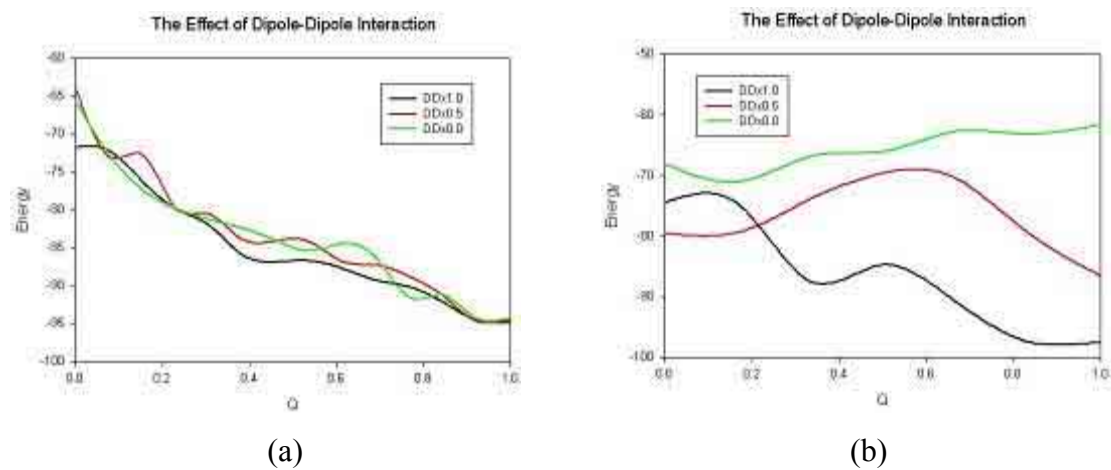

(a)                 (b)

Fig. 4.13 The effect of dipole-dipole interactions for the (a) one helix case and (b) one beta sheet case.

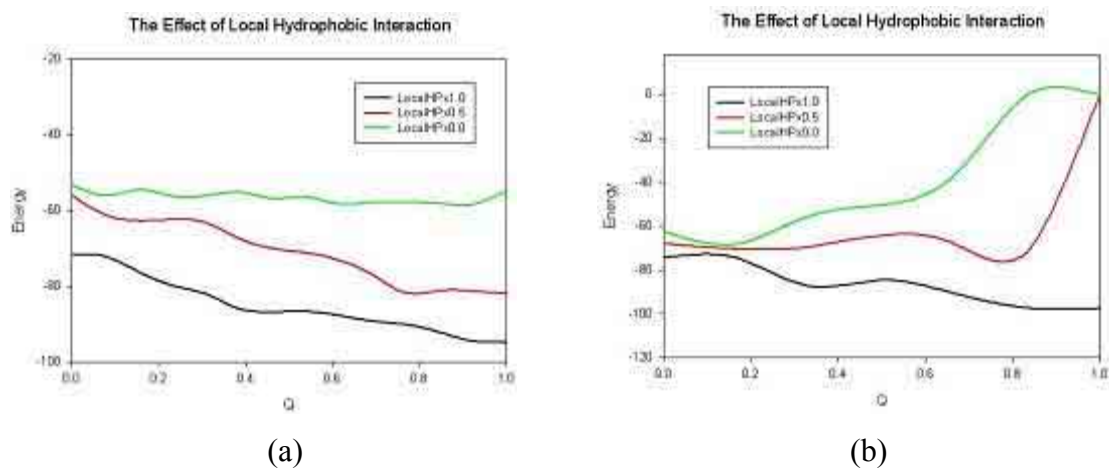

(a)                        (b)

Fig. 4.14 The effect of local hydrophobic interactions for the (a) one helix case and (b) one sheet case.

We now consider the effect of local hydrophobic interactions as shown in similar plots in Figs. 4.14 (a) and (b). Fig. 4.14 (a) shows that when the strength of local hydrophobic interactions decreases to the half of the original one, the native state for helix structure is still the same. The only difference is the landscape profile has shifted up in energy about $20.0 kcal/mol$. However, when the interactions are turned off, the native state is gone. In this case, we find that the peptide has no specific conformation for the lowest energy. On the other hand, for the beta sheet structure, when the strength is reduced, clearly, the native state is no longer preserved. Furthermore, the energy landscape profiles are very sensitive to the change. Thus, we must conclude that the local hydrophobic interactions affect both the helix and the beta sheet structures.

The upshots of the above analyses show that the two new ingredients in our model are indeed relevant for forming the secondary structures. As a final comparison, we show the effect of hydrogen bond interaction in Figs. 4.15 (a) and (b). Clearly, it shows that the hydrogen bond interaction is also essential for the secondary structures as already speculated by Pauling and has been known in many other works. Our results further indicate that the hydrogen bond interaction affects the helix more dramatically. This is because the helix structure has richer hydrogen bonds than the sheet structure does. In summary, we have demonstrated in this subsection that hydrogen bond interaction, dipole-dipole interactions, and local hydrophobic interactions all play important roles in forming the secondary structures.

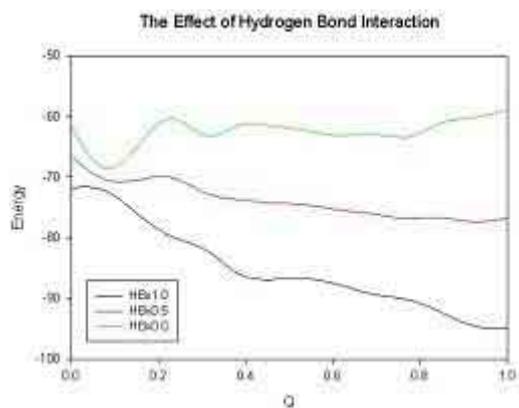 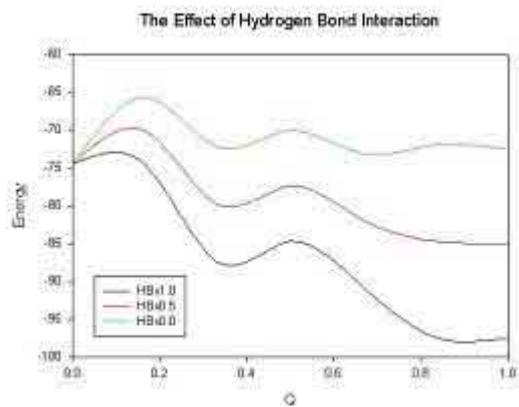

(a)                                               (b)

Fig. 4.15 The effect of hydrogen bond interaction for the (a) one helix case and (b) one beta sheet case.

## §4.2  Real Protein Peptides

The extensive analyses performed in the last section indicate that the parameters thus calibrated seem to capture the main features of simulated proteins. To further test our model and these calibrated parameters, in this section, we apply them to simulate real protein peptides whose structures are known. The PDB codes of these peptides are 1BYZ, 1DJF, 1L4X, 1NJ0, 1IBN, 1PEI, 1OEG, 3ZNF, 1PIQ, 1LYP, 1BB1, 1ZDD, 1PPT, and 1VII, respectively. These are small peptides with secondary structures that we have collected and our model has simulated with success. Some of them, such as 1VII and 3ZNF, are frequently discussed in the literature. While these proteins are small, it is by no means that being able to fold all of them is trivial. Furthermore, as their sizes have reached to the level that similar extensive analyses performed in the last section can not be done here due to consideration of the computing time, hence we will only present our simulated native states with appropriate characterization. While it may sound that this section is a collection of "benchmarks" of our model, nevertheless, we are not intending to emphasize the accuracy of our model or its capability of fold protein. Therefore, these data only serve as the purpose for demonstrating that indeed our reduced model has a good starting point.

The data for simulated native states are arranged as follows: In Table 4.4, we list basic characterization of these peptides including the sequence length, the sequences, and the main secondary structures. In Table 4.5, useful characterizations of all the folded native states are tabulated. Note that except for 3ZNF, 1PIQ, 1PPT, and 1VII, the *RMSD* (calculated for all $C^\alpha$ atoms) of all simulated native states are less than 5.0Å. In Table 4.6, the RMSD data based on different atoms are presented. Finally, Table A in appendix collects the visual forms and the contact maps for all simulated native states with a comparison to the states that are either adapted from experiments or generally are believed to be the correct native states.

Table 4.4 Basic Characterization of peptides that have been simulated successfully.

| PDB code | Seq. # | Sequences | Main structures |
|---|---|---|---|
| 1BYZ | 12 | ELLKKLLEELKG | Helix |
| 1DJF | 15 | QAPAYKKAAKKLAES | Helix |
| 1L4X | 15 | DELERAIRELAARIK | Helix |
| 1NJ0 | 16 | RKRIHIGPGRAFYTTK | Sheet |
| 1IBN | 20 | GLFGAIAGFIENGWEGMIDG | Helixes |
| 1PEI | 22 | VEEKSIDLIQKWEEKSREFIGS | Helix |
| 1OEG | 23 | PLVEDMQRQWAGLVEKVQAAVGT | Helix |
| 3ZNF | 30 | RPYHCSYCNFSFKTKGNLTKHMKSKAHSKK | Helix,Sheet |
| 1PIQ | 31 | RMKQIEDKIEEILSKQYHIENEIARIKKLIG | Helix |
| 1LYP | 32 | GLRKRLRKFRNKIKEKLKKIGQKIQGLLPKLA | Helix |
| 1BB1 | 34 | AEIAAIEYEQAAIKEEIAAIKDKIAAIKEYIAAI | Helix |
| 1ZDD | 34 | FNMQCQRRFYEALHDPN LNEEQRNAKIKSIRDDC | Two Helixes Bundle |
| 1PPT | 36 | GPSQPTYPGDDAPVEDLI RFYDNLQQYLNVVTRHRY | Helix |
| 1VII | 36 | MLSDEDFKAVFGMTRSAF ANLPLWKQQNLKKEKGLF | Helixes |
| DeGrado sequence | 54 | KELEELLKKLKELLKEGGGKELEELLK KLKELLKEGGGKELEELLKKLKELLKE | Three Helixes Bundle |

Table 4.5 "Benchmark" of RPFM: simulated native states of small real peptides. Here we characterize the native states by ground state energy (GSE), $R_G$, WAV, Q and RMSD (calculated for all $C^{\alpha}$ atoms). Note that NS, Sim, and MCsteps are abbreviations for native state, simulation results and Monte Carlo steps, respectively. The last one, DeGrado sequence [42], has no native structure to compare $R_G$, WAV, Q and RMSD. Hence, "na", which means "not available", are put in some columns.

| PDB code | GSE, kcal/mol | $R_G$, Å | | WAV, Å$^3$ | | Q | RMSD, Å | MCsteps $\overline{10^7}$ |
|---|---|---|---|---|---|---|---|---|
| | | NS | Sim | NS | Sim | | | |
| 1BYZ | −96.2 | 5.853 | 6.031 | 5774 | 6084 | 0.923077 | 1.4650 | 24.0 |
| 1DJF | −117.2 | 8.782 | 7.930 | 7436 | 7171 | 0.750000 | 2.6396 | 3.4 |
| 1L4X | −129.3 | 7.042 | 7.951 | 7332 | 7556 | 0.750000 | 2.2158 | 11.0 |
| 1NJ0 | −115.5 | 8.469 | 7.427 | 8120 | 7765 | 0.470588 | 3.5714 | 11.0 |
| 1IBN | −138.8 | 8.943 | 7.266 | 7926 | 7667 | 0.714286 | 3.7758 | 1.8 |
| 1PEI | −191.2 | 10.484 | 9.933 | 10243 | 9585 | 0.863636 | 2.2651 | 4.4 |
| 1OEG | −176.1 | 10.662 | 10.971 | 10355 | 10778 | 0.681818 | 2.7816 | 5.4 |
| 3ZNF | −234.7 | 8.439 | 9.296 | 10879 | 12510 | 0.200000 | 5.8188 | 17.0 |
| 1PIQ | −262.0 | 14.065 | 11.606 | 12420 | 12823 | 0.543478 | 6.6259 | 5.4 |
| 1LYP | −262.5 | 15.175 | 12.293 | 13928 | 13064 | 0.794118 | 2.7162 | 25.0 |
| 1BB1 | −342.5 | 14.637 | 14.973 | 11590 | 13248 | 0.893617 | 2.0816 | 20.0 |
| 1ZDD | −306.8 | 9.243 | 9.271 | 11943 | 11736 | 0.704918 | 4.5887 | 25.0 |
| 1PPT | −260.4 | 10.863 | 12.281 | 14241 | 15209 | 0.509804 | 5.4042 | 11.0 |
| 1VII | −293.8 | 8.939 | 8.922 | 11599 | 12290 | 0.541667 | 5.6783 | 8.0 |
| DeGrado sequence | −484.1 | na | 10.242 | na | 14011 | na | na | 29.1 |

Table 4.6 The *RMSD* of native states based on different atoms. Here BB indicates that the calculation is based on all backbone units, while "all" refers to calculations based on both backbone units and side-chain units (as a whole unit). Similarly, $C^\alpha$ indicates that the calculation is based on all $C^\alpha$ atoms, and so on.

| PDB code | all | BB | $C^\alpha$ | N | C | O | $C^\beta$ | $C^\alpha$ & $C^\beta$ |
|---|---|---|---|---|---|---|---|---|
| 1BYZ | 1.6172 | 1.7088 | 1.4650 | 1.0057 | 1.7712 | 2.3206 | 1.1344 | 1.3173 |
| 1DJF | 2.7592 | 2.6274 | 2.6396 | 2.6471 | 2.4364 | 2.7751 | 3.2333 | 2.9514 |
| 1L4X | 2.2656 | 2.0629 | 2.2158 | 1.9435 | 1.8783 | 2.1927 | 2.9399 | 2.6031 |
| 1NJ0 | 3.5850 | 3.4852 | 3.5714 | 3.2361 | 3.3879 | 3.7259 | 4.0097 | 3.7823 |
| 1IBN | 3.7572 | 3.7636 | 3.7758 | 3.2742 | 3.6918 | 4.2487 | 3.7207 | 3.7532 |
| 1PEI | 2.2763 | 2.0534 | 2.2651 | 2.0421 | 1.7803 | 2.0964 | 3.0374 | 2.6704 |
| 1OEG | 3.0453 | 2.9795 | 2.7816 | 2.2699 | 3.0215 | 3.6728 | 3.3185 | 3.0497 |
| 3ZNF | 5.9452 | 5.8121 | 5.8188 | 5.4934 | 5.7129 | 6.2006 | 6.4672 | 6.1460 |
| 1PIQ | 6.5949 | 6.4936 | 6.6259 | 6.7282 | 6.3018 | 6.3073 | 6.9982 | 6.8115 |
| 1LYP | 2.7266 | 2.5914 | 2.7162 | 2.3072 | 2.4894 | 2.8220 | 3.2571 | 2.9856 |
| 1BB1 | 2.0789 | 1.8787 | 2.0816 | 1.6866 | 1.7706 | 1.9508 | 2.7369 | 2.4315 |
| 1ZDD | 4.5611 | 4.1963 | 4.5887 | 4.1412 | 3.8353 | 4.1859 | 5.7948 | 5.2266 |
| 1PPT | 5.3892 | 5.3136 | 5.4042 | 4.6671 | 5.2797 | 5.8067 | 5.6941 | 5.5469 |
| 1VII | 5.7341 | 5.5093 | 5.6783 | 5.2183 | 5.3754 | 5.7483 | 6.6018 | 6.1442 |

# Chapter 5

# Conclusion and Outlook

To conclude, in this thesis, a reduced model with effective potentials that can fold both helix and sheet structures simultaneously is constructed and tested. Two important interactions, *dipole-dipole interactions* ($V_{DD}$ and $V_{DN}$) and *local hydrophobic interactions* ($V_{LocalHP}$), are introduced and are shown to be as crucial as the hydrogen bond interaction ($V_{HB}$) for forming the secondary structures. It has been demonstrated that our model is able to fold not only artificial peptides and *de novo* designed peptides but also 15 real protein peptides as well. While the water molecules are not included explicitly in this model, their effects are incorporated into several effective potentials. Without introducing any biased potentials, all testing peptides can fold to its native state in acceptable computing time. All of these testings can be best summarized in Fig. 5.1, where scatter plots of Ramachandran angles of all the simulation results are shown. The similarity of this plot to the Ramachandran plot for all proteins clearly shows the potential of our model.

By analyzing three conformation parameters, $R_G$, $Q$, and *WAV*, the energy landscapes can be mapped out in a clear way. Many previous observed folding behaviors are reproduced in our analyses. In particular, our simulation results indicate that for small artificial peptides, folding is a first-order-like transition. Furthermore, by systematically changing the relative strength of $V_{DD}$, $V_{DN}$, $V_{LocalHP}$ and $V_{HB}$, their roles for forming the secondary structures are summarized in the following figure:

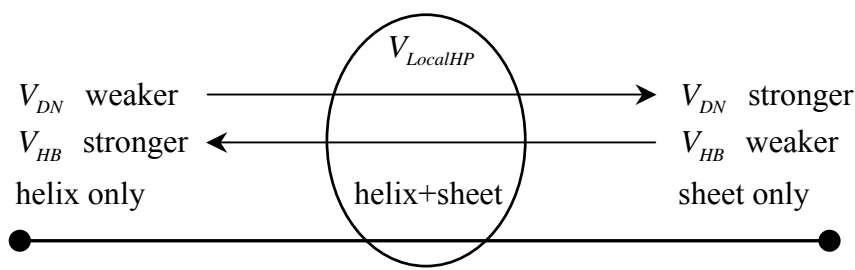

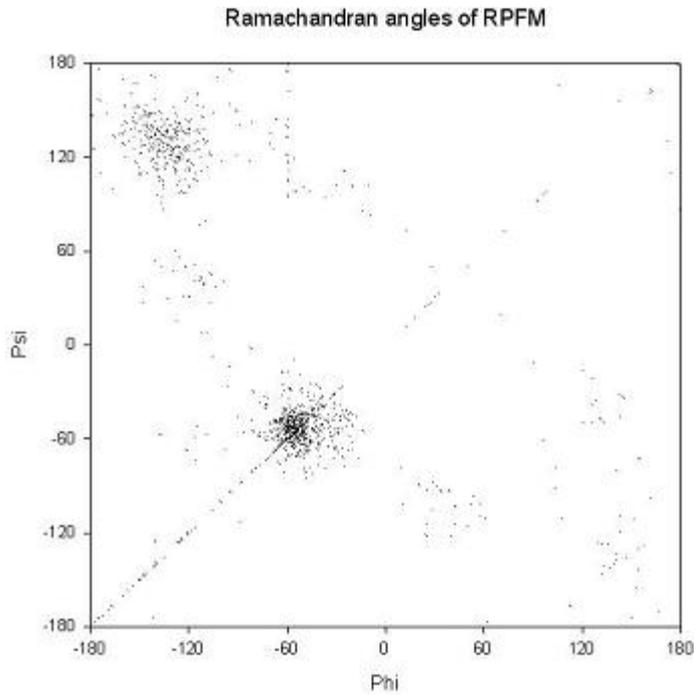

Fig. 5.1 The scatter plots of Ramachandran angles of all the simulation results in RPFM. There are two clusters. The lower one corresponds to the helix structures, while the upper one corresponds to the sheet structures.

On one hand, if the strength of $V_{HB}$ dominates, then all peptides will become helix structures. On the other hand, if the potential $V_{DN}$ dominates, then all peptides will become sheet like. Only when a subtle balance between these two interactions is hold, the helix and sheet structures can co-exist. At this point, an obvious question remains to be addressed: Since both $V_{DN}$ and $V_{HB}$ are sequence independent, for a given sequence, what determines whether it should fold into a helix or a beta sheet? This is where $V_{LocalHP}$ comes into play. We find that due to its sequence-dependent nature, $V_{LocalHP}$ is responsible for final selection for forming either a helix or a beta sheet.

After forming the secondary structures, the global hydrophobic interaction, $V_{MJ}$, is responsible for making the peptides compact so that the tertiary structures can form. Although all interactions between residues specified in $V_{MJ}$ are attractive, the difference in the interacting strength for different residue pairs assigned by the MJ matrix causes the hydrophobic residues to gain large energy when they contact to each other. This effect then makes the protein compact and enables the formation of the tertiary structure. It is important to note that the main task of the global hydrophobic interaction is to make the protein compact. This is also true for the secondary structure; especially, in the early stage of folding, it is $V_{MJ}$ that is responsible to collapse all residues into a compact space so that $V_{DD}$, $V_{DN}$, $V_{LocalHP}$ and $V_{HB}$ could begin to function, compete with each other and engineer the final conformation of the protein. In our model, $V_{MJ}$ does not play the role for deciding the secondary structure explicitly; instead, it plays a crucial role to make the formation of the secondary structure more efficient by the initial collapsing. Nevertheless, if the initial collapsing does not go in the right direction or happens too fast, the final protein structure may have no structure at all or simply form random-coil like structure. For instance, if the form of the Lennard-Jones potential is used for the $V_{MJ}$, we found that it is very easy for the peptide to get entangled in early stage before it can reach the native state. Such problems can be overcome if one uses the desolvation model to replace the Lennard-Jones potential. Obviously, $V_{MJ}$ may play an important role in forming the so-called random-coil structure.

Even though our model has clarified the roles for different potentials with promising simulation results, there are several issues that need to be pursued beyond the current work. First, as already pointed out, the precise relation of $V_{MJ}$ to the formation of random-coil needs to be pinned down. Secondly, the calibration of the energy strength is not complete yet. Only 19 peptides are demonstrated to be consistent with the current set of energy strengths. Extension to including more proteins is obviously necessary. Finally, the side-chains in our model have very simplified representations. It may be necessary to consider the internal structures with more realistic models for each side-chain. Even thought there is no definite answer for the above issues right now, the results we obtain in this thesis will serve a useful starting point for the future work.

# Bibliography

[1] Jens C. Skou. The identification of the Sodium-Potassium Pump. 1997, Nobel Lecture, http://www.nobel.se/chemistry/laureates/1997/illpres/skou.html.

[2] C. B. Anfinsen. Principles that govern the folding of protein chains. Science, 1973, **181**, 223-224.

[3] C. Levinthal. Are there pathways for protein folding? J. Chim. Phys., 1968, **65**, 44-45.

[4] C. Dobson, A. Sali, and M. Karplus. Protein folding: A perspective from theory and experiment. Angew. Chem. Int. Ed., 1998, **37**, 868-893.

[5] P. E. Leopold, M. Montal and J. N. Onuchic. Protein folding funnels – A kinetic approach to the sequence structure relationship. Proc. Nat. Acad. Sci. USA, 1992, **89**, 8721-8725.

[6] J. D. Bryngelson, J. N. Onuchic, N. D. Socci, and P. G. Wolynes. Funnels, pathways, and the energy landscape of protein folding – A synthesis. Protein: Struct. Funct. Genet., 1995, **21**, 167-195.

[7] P. G. Wolynes, J. N. Onuchic, and D. Thirumalai. Navigating the folding routes. Science, 1995, **267**, 1619-1629.

[8] V. I. Abkevich, A. M. Gutin, and E. I. Shakhnovich. Specific nucleus as the transition-state for protein-folding – Evidence from the lattice model. Biochemistry, 1994, **33**, 10026-10036.

[9] K. A. Dill, S. Bromberg, K. Z. Yue, K. M. Fiebig, D. P. Yee, P. D. Thomas, and H. S. Chan. Principles of protein folding – A perspective from simple exact models. Protein Sci., 1995, **4**, 561-602.

[10] D. K. Klimov and D. Thirumalai. Linking rates of folding in lattice models of proteins with underlying thermodynamics characteristics. J. Chem. Phys., 1998, **109**, 4119-4125.

[11] Y. Duan, L. Wang, and P. A. Kollman. The early stage of folding of villin headpiece subdomain observed in a 200-nanosecond fully solvated molecular dynamics simulation. Proc. Nat. Acad. Sci. USA, 1998, **95**, 9897-9902.

[12] Y. Duan and P. A. Kollman. Pathways to a protein folding intermediate observed in a 1-microsecond simulation in aqueous solution. Science, 1998, **282**, 5389, 740-744.

[13] H. Nymeyer, A. E. Garcia, and J. N. Onuchic. Folding funnels and frustration in off-lattice minimalist protein landscapes. Proc. Nat. Acad. Sci. USA, 1998, **95**, 5921-5928.

[14] Y. Zhou and M. Karplus. Folding thermodynamics of a model three-helix-bundle protein. Proc. Nat. Acad. Sci. USA, 1997, **94**, 14429-14432.

[15] C. Micheletti, J. R. Banavor, A. Martian, and F. Seno. Protein structures and optimal folding from a geometrical variational principle. Phys. Rev. Lett., 1999, **82**, 3372-3375.

[16] J. E. Shea, J. N. Onuchic, and C. L. Brooks III, Exploring the origins of topological frustration: Design of a minimally frustrated model of fragment B of protein A. Proc. Nat. Acad. Sci. USA, 1999, **96**, 12512-12517.

[17] Y. Zhou and M. Karplus. Interpreting the folding kinetics of helical proteins. Nature, 1999, **401**, 400-403.

[18] C. Clementi, P. A. Jennings, and J. N. Onuchic. How native-state topology affect the folding of dihydrofolate reductase and interleukin-1$\beta$. Proc. Nat. Acad. Sci. USA, 2000, **97**, 5871-5876.

[19] N. Go and H. Taketomi. Respective roles of short- and long-range interactions in protein folding. Proc. Nat. Acad. Sci. USA, 1978, **75**, 559-563.

[20] Z. Guo and D. Thirumalai. Kinetics and thermodynamics of folding of a *de Novo* designed four-helix bundle protein. J. Mol. Biol., 1996, **263**, 323-343.

[21] S. Takada, Z. L. Schulten and P. G. Wolynes. Folding dynamics with nonadditive forces: A simulation study of a designed helical protein and a random heteropolymer. J. Chem. Phys., 1999, **110**, 11616-11629.

[22] A. Irbäck, F. Sjunnesson, and S. Wallin. Three-helix-bundle protein in a Ramachandran model. Proc. Nat. Acad. Sci. USA, 2000, **97**, 13614-13618.

[23] G. Favrin, A. Irbäck, and S. Wallin. Folding of a small helical protein using hydrogen bonds and hydrophobicity forces. Protein: Struct. Funct. Genet., 2002, **47**, 99-105.

[24] J.-E. Shea, Y. D. Nochomovitz, Z. Guo, and C. L. Brooks III. Exploring the space of protein folding Hamiltonians: The balance of forces in a minimalist $\beta$-barrel model. J. Chem. Phys., 1998, **109**, 2895-2903.

[25] D. K. Klimov and D. Thirumalai. Mechanisms and kinetics of $\beta$-hairpin formation. Proc. Nat. Acad. Sci. USA, 2000, **97**, 2544-2549.

[26] S. Miyazawa and R. L. Jernigan. Residue–residue potentials with a favorable contact pair term and an unfavorable high packing density term, for simulation and threading. J. Mol. Biol., 1996, **256**, 623-644.

[27] Z.-H. Wang and H. C. Lee. Origin of the native driving force for protein folding. Phys. Rev. Lett., 2000, **84**, 574-577.

[28] C. K. Mathews, K. E. van Holde, and K. G. Ahern. Biochemistry, 3rd Ed., 2000.

[29] N. Metropolis, A.W. Rosenbluth, M.N. Rosenbluth, A.H. Teller and E. Teller, Equation of state calculations by fast computing machines. J. Chem. Phys., 1953, **21**, 1087-1092.

[30] D. P. Landau and K. Binder. A Guide to Monte Carlo Simulations in Statistical Physics, 2000.

[31] G. Humer, S. Garde, A. E. Garcia, A. Pohorille, and L. R. Pratt. An information theory model of hydrophobic interactions. Proc. Nat. Acad. Sci. USA, 1996, **93**, 8951-8955.

[32] G. Humer, S. Garde, A. E. Garcia, M. E. Paulaitis, and L. R. Pratt. The pressure dependence of hydrophobic interactions is consistent with the observed pressure denaturation of proteins. Proc. Nat. Acad. Sci. USA, 1998, **95**, 1552-1555.

[33] N. Hillson, J. N. Onuchic, and A. E. Garcia. Pressure-induced protein-folding / unfolding kinetics. Proc. Nat. Acad. Sci. USA, 1999, **96**, 14848-14853.

[34] M. S. Cheung, A. E. Garcia, and J. N. Onuchic. Protein folding mediated by solvation: Water expulsion and formation of the hydrophobic core occur after the structural collapse. Proc. Nat. Acad. Sci. USA, 2002, **99**, 685-690.

[35] G. Solomons and C. Fryhle. Organic Chemistry, 7th Ed., 2000.

[36] A. M. Lesk. Introduction to Protein Architecture, 2001.

[37] A. V. Finkelstein and O. B. Ptitsyn. Protein Physics A Course of Lectures, 2002.

[38] C. Branden an J. Tooze. Introduction to Protein Structure, 2nd Ed., 1999.

[39] The standard form with the threefold symmetry is $\frac{\varepsilon}{2}\sum_i (1+\cos 3\theta_i)$, where $\theta_i = \phi_i$ or $\psi_i$.

[40] M. Daune. Molecular Biophysics Structures in Motion, 1993.

[41] http://www.umass.edu/microbio/rasmol/

[42] L. Regan and W. F. DeGrado. Characterization of a helical protein designed from first principles. Science, 1988, **241**, 976-978.